\documentclass[journal]{IEEEtran}

\usepackage{cite}
\usepackage{amsmath,amssymb,amsfonts}
\usepackage{algorithmic}
\usepackage{graphicx}
\usepackage{textcomp}
\usepackage{xcolor}
\usepackage{cuted}
\usepackage{epstopdf}
\usepackage{stfloats}
\usepackage{algorithm}
\usepackage{enumerate}
\usepackage{caption}
\usepackage{comment}

\usepackage[caption=false,font=footnotesize,labelfont=rm,textfont=rm]{subfig}

\usepackage{svg}

\usepackage{hyperref}
\usepackage{amsthm}
\theoremstyle{plain}

\begin{document}

%Throughput Fairness for Movable Intelligent Surface–Enabled Periodic IoT Networks: Static Phase-Shift Design and User Scheduling
%Movable Intelligent Surface (MIS)-Enabled Wireless Communications: When Static Phase Element Meets Tunable Relative Position
\title{Wireless Sensing with Movable Intelligent Surface}
\author{Ziyuan~Zheng, Qingqing~Wu, Yanze Zhu, Wen Chen, Ying Gao, and Honghao Wang \vspace{-12pt}
\thanks{The authors are with the Department of Electronic Engineering, Shanghai Jiao Tong University, 200240, China (e-mail: \{zhengziyuan2024, qingqingwu, yanzezhu, wenchen, yinggao\}@sjtu.edu.cn); mc25018@um.edu.mo).}
}% <-this % stops a space

\markboth{}%
{Shell \MakeLowercase{\textit{et al.}}: Bare Demo of IEEEtran.cls for IEEE Journals}

\maketitle


\begin{abstract}
Future wireless networks are envisioned to deliver not only gigabit communications but also ubiquitous sensing. Reconfigurable intelligent surfaces (RISs) have emerged to reshape radio propagation, recently showing considerable promise for wireless sensing. Still, their per-element electronic tuning incurs prohibitive hardware cost and power consumption. Motivated by the concept of fluid antenna system (FAS), this paper introduces a low-cost movable intelligent surface (MIS) for wireless sensing, which replaces element-wise electronic phase tuning with panel-wise mechanical reconfiguration. The MIS stacks a large fixed and a smaller movable pre-phased metasurface layers, whose differential position shifts synthesize distinct composite phase patterns, enabling multiple beam patterns for multi-target detection. We characterize a MIS-enabled multi-hop echo signal model with multi-target interference and then formulate a worst-case sensing signal-to-interference-plus-noise ratio (SINR) maximization problem that jointly designs MIS phase shifts and schedules MS2’s position. A Riemannian Augmented Lagrangian Method (RALM)-based algorithm is developed to solve the formulated mixed-integer non-convex problem. We also derive a heuristic MIS beam steering design with closed-form phase distribution and position scheduling. Simulations validate MIS’s beam pattern reconfiguration capability, show that the RALM-based scheme significantly outperforms the closed-form scheme in improving sensing SINR, and uncover a gain–diversity trade-off in beam patterns that informs the optimal choice of MIS configuration.  
\end{abstract}

% Note that keywords are not normally used for peer-reviewed papers.
\begin{IEEEkeywords}
Movable intelligent surface (MIS), fluid antenna system (FAS), mechanical reconfigurability, differential position shifting, beam pattern scheduling.
\end{IEEEkeywords}

% \fi
%
% For peer review papers, this IEEEtran command inserts a page break and
\IEEEpeerreviewmaketitle

\vspace{-6pt}
\section{Introduction}

The next generation wireless infrastructure envisions integrated sensing and communication (ISAC) functionality for applications such as environment monitoring, vehicle localization, immersive interaction, and industrial automation [1]. Realizing ISAC requires shaping radio signals to illuminate targets and capture their reflections, often under stringent energy and cost constraints [2]. However, conventional wireless architectures with uncontrollable propagation environments struggle to provide ubiquitous sensing capabilities, and achieving high cost-effectiveness poses even greater challenges [3]. This gap has catalyzed interest in technologies that reconfigure the propagation environment itself to jointly enhance the coverage and accuracy of wireless sensing [4]. Among these, the reconfigurable intelligent surface (RIS), a programmable metasurface (MS) composed of massive phase-shift elements that can intelligently redirect incident signals, has emerged as a promising candidate [5]. Coordinated control of phase elements enables RIS to shape reflected or transmitted electromagnetic waves, delivering passive beamforming gains without additional active radio frequency (RF) chains or power-hungry amplifiers [6]. Extensive studies reveal its benefits for throughput, coverage, and energy efficiency [7]. More recently, in addition to communication, RIS has also been shown to improve wireless sensing, including target detection and localization scenarios, by redirecting illumination toward non-line-of-sight (NLoS) targets or acting as a passive reflector from favorable angles that enriches spatial diversity [8]. These advances underscore the considerable promise of intelligent surfaces for wireless sensing in future networks. 

Despite this promise, fully dynamic, element-wise tunable RISs face two key hardware bottlenecks that restrict their large-scale deployment in practical wireless sensing applications [9]. The first is \textit{dense cabling}. Every tunable unit requires bias lines, clock signals, and often high-speed digital interfaces. When the number of dynamic units reaches hundreds or thousands, printed circuit routing, connectors, and shielding lead to a bulky hardware stack [10]. Parasitic coupling at millimeter wave frequencies further degrades phase accuracy, while sheer pin count inflates packaging cost and threatens long-term reliability as the aperture grows. The second is \textit{power consumption.} Digital analog converters, voltage regulators, and phase-shifting components dissipate milliwatts of standby power even when idle. Therefore, a dynamic RIS of elements in thousands of orders idles at several watts, undermining the popular selling point of nearly zero power consumption, which also challenges thermal budgets for outdoor or lightweight deployment platforms, in contradiction to the low-cost appeal of passive surfaces [11]. Recent advanced RIS architectures proposed to increase functionality or push performance boundaries, such as active RIS [12], stacked intelligent MS (SIM) [13], simultaneous-transmitting-and-reflecting (STAR)-RIS [14], and beyond-diagonal (BD)-RIS [15], further exacerbate these pain points by adding extra active chains or switching networks. 

In contrast, one alternative approach to simplify the implementation of RIS is to forego continuous reconfiguration and use static MSs (SMS) that have fixed phase patterns [16]. 
An SMS can be pre-designed for a specific task or coverage area and then left unchanged, drastically reducing hardware complexity by eliminating electronics and tuning circuitry [17]. For example, a reflecting SMS can use its optimized phase pattern to extend coverage to a certain zone or to create a permanent reflecting wall to redirect signals to blind spots [18]. However, SMSs generally support only a single beam pattern, or a limited set of patterns, determined at design time, lacking the adaptability required for multi-direction or time-varying wireless applications [19]. In a sensing context, an SMS might be able to assist with a predetermined sensing direction or area but would be unable to re-target its beam to track moving objects or to sequentially scan different areas of interest [20]. Thus, while SMS are appealing due to their simplicity and cost-effectiveness for static wireless applications, they are inadequate when dynamic beam steering or multi-directional sensing is required [21]. Consequently, dynamic RIS and SMS present a trade-off between performance and cost: dynamic RISs offer fine beam control but require dense cabling, continuous power consumption, and substantial signaling overhead, whereas low-cost static surfaces require no electronics but are limited to a single beam pattern. 

This disparity leaves a practical gap in quasi-static wireless applications, such as industrial Internet of Things (IoT), sensor networks, and environment monitoring, where channels are stable and user demands change only occasionally or periodically [22]; neither extreme, that is, dynamic RIS and static surface, is sufficiently economical or flexible. To bridge this gap, recent work explores architectures that balance flexibility and complexity, notably mechanically reconfigurable MSs: instead of per-element electronic tuning, the entire surface, or a sublayer, is shifted, rotated, or morphed to steer beams [23],[24]. This concept leverages the fact that changing the geometry alters the aggregate radiation pattern, enabling beam control with minimal electronics [25]. Early demonstrations include moiré MS, where rotating two patterned layers produces tunable phase distribution and beam steering without per-element phase shifters [26], and kirigami MS, whose stretch and bend deformations re-point or re-focus radio energy [27]. This paradigm can replace or augment electronic tuning, reducing dense cabling and standby power while retaining adaptability. Although mechanical adjustment may be slower than electronic control, many slowly varying sensing tasks, such as periodic scanning, area surveillance, and monitoring environments [28], do not strictly necessitate high-speed reconfiguration with sub-millisecond agility; beams may only need to be switched or moved on the order of milliseconds to seconds, which is feasible with mechanical reconfigurability.

Building on this paradigm, we propose the movable intelligent surface (MIS) for wireless sensing. MIS is a two-layer closely stacked architecture comprising a large primary layer, denoted by MS1, and a smaller secondary layer, denoted by MS2, that slides discretely within the area of MS1, each with a pre-designed static phase pattern. With differential position shifting, each mutual displacement creates a distinct composite phase distribution, thereby synthesizing different radiation patterns without altering individual element phases. Conceptually, one can design the static phase patterns of MS1 and MS2 such that each offset of MS2 yields a beam directed towards a distinct angle or target region of interest. By mechanically repositioning MS2 into a schedule of discrete positions, the MIS can scan across these pre-defined beam directions. Crucially, this beam switching is achieved without any per-element electronic tuning, which greatly reduces the hardware complexity compared to a fully tunable RIS, while preserving the favored beamforming adaptability absent in single-layer static surfaces. 

The main contributions of this paper are listed below.
\begin{itemize}
\item First, we introduce the MIS, composed of a fixed MS1 and a movable MS2 with static phase elements, for low-cost wireless sensing. Based on MIS architecture, we propose the differential position shifting mechanism for beam pattern reconfiguration without per-element tuning. 
\item Second, we characterize the MIS-enabled multi-hop echo signal model with multi-target interference. Following the signal modeling, we formulate a new optimization problem to design MIS phase shifts and schedule MS2's position for worst-case sensing signal-to-interference-plus-noise ratio (SINR) maximization.
\item Third, we develop an efficient algorithm based on the Riemannian Augmented Lagrangian Method (RALM) to solve the formulated mixed-integer non-convex program. We recast the original problem into a smooth manifold-constraint formulation and solve it with RALM on a product manifold in an iterative manner.
\item Fourth, we derive a heuristic MIS beam steering design with closed-form phase distribution and position-shifting expressions. This scheme achieves two-dimensional beam steering for wireless sensing, while it does not explicitly address interference, serving as a general benchmark. 
\item Finally, numerical results validate the beam pattern reconfiguration capability of the MIS for wireless sensing, highlight the superiority of the RALM-based scheme over the closed-form scheme for interference suppression, reveal the trade-off between beam pattern gain and diversity, and draw insight to the optimal MIS configuration.
\end{itemize}
The remainder of this paper is organized as follows. Section II introduces the MIS and its reconfiguration mechanism. Section III presents the system and signal model and formulates the minimum sensing SINR maximization problem. Section IV details the RALM-based algorithm. Section V proposes the closed-form heuristic design. Section VI discusses the numerical results, and Section VII concludes the paper.

\section{MIS Architecture and Modeling}

As illustrated in Fig. 1, we consider an MIS-aided wireless sensing system, where a transmissive MIS is deployed to assist an $L$-antenna base station (BS) in detecting objects within designated target directions. Due to obstructions in the direct BS-target paths, the MIS refracts signals from the BS toward the target region, creating virtual line-of-sight (LoS) channels to facilitate reliable sensing. This section describes the proposed MIS architecture and the dynamic beamforming mechanism enabled by its movable structure, then establishes a representation of the composite MIS phase shift vector.

\vspace{-6pt}
\subsection{MIS Architecture and Beamforming}
\setlength{\abovecaptionskip}{0pt}
\begin{figure}[t]
    \centering
    \includegraphics[width=2.8in]{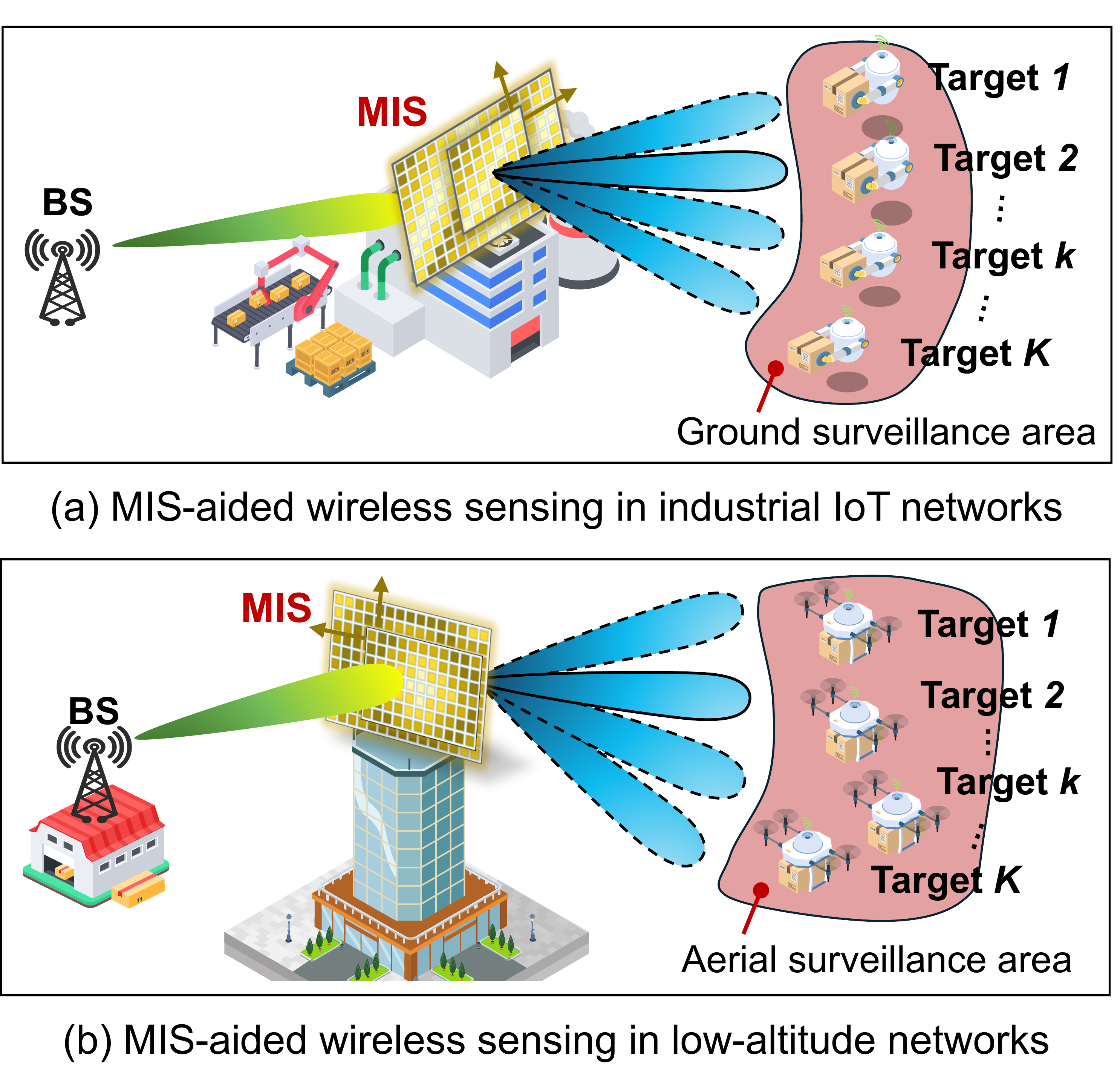}
    \captionsetup{font=small}
    \caption{Illustration of MIS-aided wireless sensing for  area surveillance in ground industrial IoT and aerial low-altitude networks.} 
    \label{fig:system_model}
    \vspace{-12pt}
\end{figure}

\begin{comment}
\setlength{\abovecaptionskip}{6pt}
\begin{figure}[t]
    \centering
    \includegraphics[width=3in]{system2.pdf}
    \captionsetup{font=small}
    \caption{An example of the MIS beam pattern switching mechanism for target detection under a max-min signal-to-interference-plus-noise (SINR) rule. By sliding MS2 over MS1, each with pre-designed static phase shifts, the MIS creates multiple beam patterns for designated target directions without requiring electronic tuning.} 
    \label{fig:system_model}
    \vspace{-12pt}
\end{figure}
\end{comment}

%\setlength{\abovecaptionskip}{0pt}

The proposed MIS comprises two transmissive MSs, stacked closely: a larger fixed-position MS1 and a smaller movable MS2. Both MS layers adopt pre-designed static phase shifts tailored to specific sensing tasks. MS1 is structured as a uniform planar array (UPA) with $M=M_r \times M_c$ elements spaced $d_{\text{MIS}}$, indexed as $m = (m_r - 1) M_c + m_c$, where $m_r \in \{1,\dots,M_r\}$ and $m_c \in \{1,\dots,M_c\}$. Similarly, MS2, consisting of $N=N_r \times N_c$ elements, is capable of moving within the boundary of MS1, and its elements can be indexed as $n = (n_r - 1) N_c + n_c$, where $n_r \in \{1,\dots,N_r\}$ and $n_c \in \{1,\dots,N_c\}$. The static phase shifts of MS1 and MS2 are represented by diagonal matrices $\boldsymbol{\varPhi}=\text{diag}(\boldsymbol{\phi})\in\mathbb{C}^{M\times M}$ and $\boldsymbol{\varTheta}=\text{diag}(\boldsymbol{\theta})\in\mathbb{C}^{N\times N}$, respectively, with $\boldsymbol{\phi}=[e^{j\phi_1},\dots,e^{j\phi_M}]^T$ and $\boldsymbol{\theta}=[e^{j\theta_1},\dots,e^{j\theta_N}]^T$, where $\phi_m,\theta_n\in[0,2\pi]$ denote the static phase shifts of each element.

Unlike conventional RISs that require electronic element-wise phase tuning, the MIS switches beam patterns through mechanical repositioning of MS2 relative to MS1. Termed differential position shifting, this mechanism enables MS2 to overlap different subsets of MS1 elements, thereby creating distinct composite phase patterns without electronic phase adjustments. Enabling by differential position shifting mechanism, the MIS offers beam pattern switching suitable for sensing tasks where periodic beam updates suffice, reducing complexity and cost significantly. The available beam patterns from the MIS can be enumerated as $U = U_r \times U_c$, where $U_r = M_r - N_r + 1$ and $U_c = M_c - N_c + 1$ represent the maximum index of discrete shifting positions along rows and columns, respectively. Thus, each beam pattern $u\in\mathcal{U}=\{1,\dots,U\}$ corresponds uniquely to an overlap position $(u_r,u_c)$, indexed as $u = (u_r - 1)\times U_c + u_c$.

\vspace{-9pt}
\subsection{MIS Equivalent Phase Shift Vectors}
The BS transmitted signal propagates sequentially through MS1 and MS2, or vice versa, and signals may transmit through either the overlapped regions of both layers or the non-overlapping part of MS1 alone. The close stacking of MS layers allows diffraction impacts in adjacent elements to remain negligible, preventing significant signal leakage or wavefront divergence. Therefore,  when MS2 is shifted to the position $u$, we define the equivalent phase shift vector as
\vspace{-3pt}
\begin{equation}
\boldsymbol{\bar{\theta}}_{u}= \boldsymbol{S}_u \boldsymbol{\theta} + \boldsymbol{e}_u \in \mathbb{C}^{M\times 1}, \forall u \in \mathcal{U}, 
\end{equation}
where $\boldsymbol{S}_u \in \{0,1\}^{M \times N}$ is a binary selection matrix, and $\boldsymbol{e}_u\in\{0,1\}^{M\times 1}$ is a binary padding vector defined as: $\left[ \boldsymbol{S}_u \right] _{m,n}=1$ and $\left[ \boldsymbol{e}_u \right]_m=0$ if $n$-th element of MS2 is located on $m$-th element of MS1; $\left[ \boldsymbol{S}_u \right]_{m,n}=0$ and $\left[ \boldsymbol{e}_u \right]_m=1$ otherwise. Thus, $\boldsymbol{S}_u$ maps MS2 elements to the corresponding overlapped positions of MS1, while $\boldsymbol{e}_u$ assigns virtual MS2 elements with unit amplitude and zero phase to the MS1 elements that are not overlapped by MS2, collectively characterizing the MIS overlapping pattern. 
This compact representation captures the interaction between static phase elements with dynamic geometry, aligning the effective dimensions across MS layers for unified MIS phase shift vector modeling.

\subsection{Channel Model}
To draw insights into the MIS beam steering mechanism for wireless sensing, we consider a deterministic LoS-dominated channel model. The BS-MIS channel $\boldsymbol{R}$ is modeled as: $\boldsymbol{R} = \mathbf{a}_{\text{MIS}}\mathbf{a}_{\text{BS}}^T \in \mathbb{C}^{M \times L}$, where $\mathbf{a}_{\text{MIS}} \in \mathbb{C}^{M \times 1}$ and $\mathbf{a}_{\text{BS}} \in \mathbb{C}^{L \times 1}$ represent the MIS and BS array responses, respectively. Similarly, the MIS-to-target channel vector $\mathbf{a}_k \in \mathbb{C}^{M \times 1}$ incorporates the LoS components from the MIS elements to the $k$-th target, with array response vectors defined as
\vspace{-3pt}
\begin{subequations}
\begin{align}
&\!\! \mathbf{a}_{\text{MIS}}\left( \vartheta _{\text{MIS}},\psi _{\text{MIS}} \right) \nonumber
\\
&\!\!=[1,e^{j \frac{2\pi d_{\text{MIS}}}{\lambda}\left( m_r\cos \left( \vartheta _{\text{MIS}} \right) \sin \left( \psi _{\text{MIS}} \right) +m_c\sin \left( \vartheta _{\text{MIS}} \right) \sin \left( \psi _{\text{MIS}} \right) \right)},\dots, \nonumber
\\
&\!\!\,\, e^{j \frac{2\pi d_{\text{MIS}}}{\lambda}\left(\! \left( M_r\!-\!1 \right)\! \cos \left( \vartheta _{\text{MIS}} \right) \sin \left( \psi_{\text{MIS}} \right) +\left( M_c\!-\!1 \right) \sin \left( \vartheta_{\text{MIS}} \right) \!\sin \left( \psi_{\text{MIS}} \right) \!\right)}]^T\!\!,\!\!\!\!
\\
&\!\!\mathbf{a}_{\text{BS}}\left( \vartheta _{\text{BS}},\psi _{\text{BS}} \right)
\nonumber
\\
&\!\!=[1,e^{j \frac{2\pi d_{\text{BS}}}{\lambda}\left( n_r\cos \left( \vartheta _{\text{BS}} \right) \sin \left( \psi _{\text{BS}} \right) +n_c\sin \left( \vartheta _{\text{BS}} \right) \sin \left( \psi _{\text{BS}} \right) \right)},\dots , \nonumber
\\
&\!\!\,\,e^{j \frac{2\pi d_{\text{BS}}}{\lambda}\left( \left( L_r\!-\!1 \right) \cos \left( \vartheta _{\text{BS}} \right) \sin \left( \psi _{\text{BS}} \right) +\left( L_c\!-\!1 \right) \sin \left( \vartheta _{\text{BS}} \right) \sin \left( \psi _{\text{BS}} \right) \right)}]^T\!,\!\! 
\\
&\!\!\mathbf{a}_k\left( \vartheta _k,\psi _k \right) \nonumber
\\
&\!\!=[1,e^{j \frac{2\pi d_{\text{MIS}}}{\lambda}\left( n_r\cos \left( \vartheta _k \right) \sin \left( \psi _k \right) +n_c\sin \left( \vartheta _k \right) \sin \left( \psi _k \right) \right)},\dots , \nonumber
\\
&\!\!\,\,e^{j \frac{2\pi d_{\text{MIS}}}{\lambda}\left( \left( L_r-1 \right) \cos \left( \vartheta _k \right) \sin \left( \psi _k \right) +\left( L_c-1 \right) \sin \left( \vartheta _k \right) \sin \left( \psi _k \right) \right)}]^T,
\end{align}    
\end{subequations}
where $d_{\text{BS}}$ denotes the BS antenna spacing and $(\vartheta_{\text{BS}}, \psi_{\text{BS}})$, $(\vartheta_{\text{MIS}}, \psi_{\text{MIS}})$, and $(\vartheta_k, \psi_k)$ represent the azimuth and elevation angles of departure (AoD) for the BS, the azimuth and elevation angles of arrival (AoA) for the MIS, and the azimuth and elevation AoA for the direction of $k$-th target, respectively.

\section{MIS Sensing Model and Problem Formulation}
In the following, we present the MIS-enabled wireless sensing model, where the MIS leverages its finite beam pattern set $\mathcal{U}$ to detect $K$ candidate target directions in a time-division manner, and formulate a new optimization problem.

\subsection{MIS-Enabled Wireless Sensing Model}
Let $T_p$ denote the number of pulse repetition intervals (PRIs) per dwell time, BS sensing beams are modulated by signals $\boldsymbol{s}_k = [s_k[1], \dots, s_k[T_p]]^T, \forall k \in \mathcal{K}$. Whether a target exists in the direction $k$ is determined by detecting the echo corresponding to $\boldsymbol{s}_k$, which is received in BS after a multi‑hop target reflection (i.e., BS-MIS-target-MIS-BS). During PRI $t$, if to probe a desired direction $k$, i.e., $(\vartheta_k,\varphi_k)$, BS transmits a complex baseband waveform $\boldsymbol{w}_{k,u}s_k[t]$ that cooperates with the MIS beam pattern $u$, where $\|\boldsymbol{w}_{k,u}\|^2\leq P$, with the maximum transmit power $P$, and $|s_k[t]|^2=1$. 
The BS received signal in the time interval $t$ is given by
\begin{align}
\!\!\!\boldsymbol{y}_{\!k,u}[t] \! \!=\!\!\sum_{\!k\in \mathcal{K}}{\!\boldsymbol{R}^T\boldsymbol{\varPhi }^T\boldsymbol{\bar{\varTheta}}_{u}^{T}\!\mathbf{a}_k\beta _k\mathbf{a}_{k}^{H}\boldsymbol{\bar{\varTheta}}_u\boldsymbol{\varPhi\! Rw}_{k,u}s_k[t\!-\!\tau _k]}\!+\!\boldsymbol{n}[t], \!\!\!
\end{align}
where $\beta_k$ and $\tau_k$ are the reflection coefficient multiplied by path loss and propagation delay of the round-trip signals reflected from $k$-th target, respectively, and $\boldsymbol{n}[t]$ is additive white Gaussian noise (AWGN) with covariance matrix $\sigma ^2\boldsymbol{I}_M$. 
The beam pattern gain in the target direction $\{\varphi_k,\vartheta_k \}$ is
\begin{subequations}
\begin{align}
\mathcal{P}_{k,u}&=\mathbb{E}\left[ \left| \beta_k\mathbf{a}_{k}^{H}\boldsymbol{\bar{\varTheta}}_u\boldsymbol{\varPhi Rw}_{k,u}s_k\left[ t \right] \right|^2 \right] 
\\
&=\beta_k^2\mathbf{a}_{k}^{H}\boldsymbol{\bar{\varTheta}}_u\boldsymbol{\varPhi Rw}_{k,u}\boldsymbol{w}_{k,u}^{H}\boldsymbol{R}^H\boldsymbol{\varPhi }^H\boldsymbol{\bar{\varTheta}}_{u}^{H}\mathbf{a}_{k},
\end{align}
\end{subequations}
while the beam pattern gain of the sensing beam intended for target $k$ towards the other targets is considered to cause undesired multi-target interference introduced as
\begin{subequations}
\begin{align}
\mathcal{I}_{k,u}=&\sum_{k'\in \mathcal{K}\backslash k}{\mathbb{E}\left[ \left| \beta_{k'}\mathbf{a}_{k'}^{H}\boldsymbol{\bar{\varTheta}}_u\boldsymbol{\varPhi Rw}_{k,u}s_k\left[ t-\tau _{k'} \right] \right|^2 \right]}
\\
=&\sum_{k'\in \mathcal{K}\backslash k}{\beta_{k'}^2\mathbf{a}_{k'}^{H}\boldsymbol{\bar{\varTheta}}_u\boldsymbol{\varPhi Rw}_{k,u}\boldsymbol{w}_{k,u}^{H}\boldsymbol{R}^H\boldsymbol{\varPhi }^H\boldsymbol{\bar{\varTheta}}_{u}^{H}\mathbf{a}_{k'}}.
\end{align}
\end{subequations}
This signal energy received from undesired directions, induced by power leakage, may result in poor detection performance.

With a receiver filter $\boldsymbol{f}_{k,u}^{H}$ for $k$-th target, the combined signal in the interval $t$ is $y_k\left[ t \right]=\boldsymbol{f}_{k,u}^{H}\boldsymbol{y}_k\left[ t \right]$. Due to the rank-1 characteristic of the LoS BS-MIS channel matrix $\boldsymbol{R}=\mathbf{a}_{\text{MIS}}\mathbf{a}_{\text{BS}}^T$, the matched filter $\boldsymbol{f}_{k,u}^{H}=\mathbf{a}_{\text{BS}}^{H}$ and the maximum ratio transmission beamforming $\boldsymbol{w}_{k,u}=P\mathbf{a}_{\text{BS}}^{*}$ maximize the echo signal power from $k$-th target. Accordingly, we have
\begin{align}
\tilde{y}_{k,u}[t] =&\sum_{k\in \mathcal{K}}PL^2\beta _{k}^{}\mathbf{a}_{\text{MIS}}^{T}\boldsymbol{\varPhi }^T\boldsymbol{\bar{\varTheta}}_{u}^{T}\mathbf{a}_{k}\mathbf{a}_{k}^{T}\boldsymbol{\bar{\varTheta}}_u\boldsymbol{\varPhi }\mathbf{a}_{\text{MIS}}s_k[t-\tau _k] \nonumber
\\
&+P\mathbf{a}_{\text{BS}}^{H}\boldsymbol{n}[t]. 
\end{align}
By multiplying $\tilde{\boldsymbol{y}}_{k,u}$ by $\boldsymbol{s}_k^H$, the received echo signal $z_{k,u}$ is
\begin{subequations}
\begin{align}
z_{k,u}&=\boldsymbol{s}_{k}^{H}\tilde{\boldsymbol{y}}_{k,u} =\alpha _{k,u}\boldsymbol{s}_{k}^{H}\boldsymbol{s}_k\!+\!\sum_{k'\in \mathcal{K}\backslash k}{\alpha _{k',u}\boldsymbol{s}_{k}^{H}\boldsymbol{s}_k}\!+\!\bar{n}_k,
\\
\alpha _{k',u}&=PL^2\beta _{k'}^{}\mathbf{a}_{\text{MIS}}^{T}\boldsymbol{\varPhi }^T\boldsymbol{\bar{\varTheta}}_{u}^{T}\mathbf{a}_{k'}^{}\mathbf{a}_{k'}^{T}\boldsymbol{\bar{\varTheta}}_u\boldsymbol{\varPhi }\mathbf{a}_{\text{MIS}},
\\
\bar{n}_k&=P\mathbf{a}_{\text{BS}}^{H}\sum_{t=1}^{T_p}{s_k\left[ t \right] \boldsymbol{n}\left[ t+\tau _k \right]}.
\end{align}
\end{subequations}
The target detection problem can then be formulated as a binary hypothesis test based on (7a), given as follows for the target $k$:
$\mathcal{H}_{k}^{0}$: No target in the $k$-th direction;
$\mathcal{H}_{k}^{1}$: A target exists in the $k$-th direction.
The detection model can be described by the following hypothesis testing problem.
\begin{align}
z_{k,u}=
\begin{cases}
	\mathcal{H}_{k}^{0}:\sum_{k'\in \mathcal{K}\backslash k}{\alpha_{k',u}\boldsymbol{s}_{k}^{H}\boldsymbol{s}_k}+\bar{n}_k
    \\
	\mathcal{H}_{k}^{1}:\alpha_{k,u}\boldsymbol{s}_{k}^{H}\boldsymbol{s}_k+\sum_{k'\in \mathcal{K}\backslash k}{\alpha_{k',u}\boldsymbol{s}_{k}^{H}\boldsymbol{s}_k}+\bar{n}_k 
    \\
\end{cases}
\!\!\!.\!\!\!
\end{align}
Then, the optimal detector is $E=\left| z_{k,u} \right|^{2}\underset{\mathcal{H}_{k}^{0}}{\overset{\mathcal{H}_{k}^{1}}{\lessgtr}}\mu$ with Gaussian distributed $\bar{n}_k$, where the decision threshold $\mu$ is set to meet the desired false alarm rate. The detection reliability increases with increasing $\left| \alpha _{k,u} \right|$ and decreasing interference $\left| \alpha _{k',u} \right|$. Therefore, the beam pattern gains of the dedicated beams in the respective target directions should be maximized while limiting the power leakage in the directions of the other targets.

\vspace{-6pt}
\subsection{Problem Formulation}
According to (8), the instantaneous sensing SINR for direction $k$ under MIS beam pattern $u$ is thus expressed as
\begin{align}
\text{SINR}_{k,u}=\frac{\beta _{k}^{2}\left( \boldsymbol{v}_{u}^{H}\boldsymbol{G}_k\boldsymbol{v}_u \right) ^2}{\sum_{i\in \mathcal{K},i\ne k}{\beta _{i}^{2}\left( \boldsymbol{v}_{u}^{H}\boldsymbol{G}_i\boldsymbol{v}_u \right) ^2}+\frac{\sigma _{k}^{2}}{PL^2}},
\end{align}
where we denote the effective passive beamforming vector $\boldsymbol{v}_u$ and the positive semi-definite matrix $\boldsymbol{G}_k$ written as 
\begin{align}
\boldsymbol{v}_u&\triangleq \boldsymbol{\bar{\theta}}_{u}^{}\odot \boldsymbol{\phi}=\mathrm{diag}\left( \boldsymbol{S}_u\boldsymbol{\theta }+\boldsymbol{e}_u \right) \boldsymbol{\phi },
\\
\boldsymbol{G}_k&\triangleq \mathrm{diag}\left( \mathbf{a}_{\text{MIS}}^{T} \right) \mathbf{a}_{k}^{}\mathbf{a}_{k}^{T}\mathrm{diag}\left( \mathbf{a}_{\text{MIS}} \right) .
\end{align}

The reliable target detection problem to maximize the minimum SINR among all targets is formulated as follows
\vspace{-6pt}
\begin{subequations}
\begin{align}
\text{(P1)}:\underset{\boldsymbol{\phi },\boldsymbol{\theta }}{\max}&\,\,\underset{k}{\min}\,\,\big\{ \underset{u}{\text{arg}\max}\,\,\left\{ \text{SINR}_{k,u} \right\} \big\} 
\\
\text{s.t.}  \,\, &\left| \phi _m \right|=1, \forall m\in \mathcal{M},
\\
&    \left| \theta _n \right|=1, \forall n\in \mathcal{N},
\end{align}
\end{subequations}
where (12b) and (12c) enforce unit-modulus phase elements of MIS.
The MIS switches among the available beam patterns to detect different targets via mechanical reconfiguration. Let the binary variable $\xi_{k,u}\in \{0,1\}$ indicate whether the direction $k$ is sensed by the MIS beam pattern $u$ in its scheduled time slot, where only one MIS beam pattern can be activated. The time division operation implies $\sum_{u\in\mathcal{U}}\xi_{k,u}=1$ for each $k$, that is, each target is assigned exactly one optimal beam pattern that maximizes its sensing SINR. Consequently, the minimum SINR maximization problem (P1) is equivalently recast as
\begin{subequations}
\begin{align}
\text{(P2)}:&\underset{\boldsymbol{\phi},\boldsymbol{\theta},\left\{\xi_{k,u}\right\}}{\max}\,\, \eta 
\\
\text{s.t.}\,\, &  \sum_{u\in \mathcal{U}}{\!\xi_{k,u}\frac{\beta _{k}^{2}\left( \boldsymbol{v}_{u}^{H}\boldsymbol{G}_k\boldsymbol{v}_u \right) ^2}{\sum_{k'\in \mathcal{K}\backslash k}{\beta _{k'}^{2}\left( \boldsymbol{v}_{u}^{H}\boldsymbol{G}_{k'}\boldsymbol{v}_u \right) ^2} \!+\!\frac{\sigma _{k}^{2}}{PL^2}}} \!\ge \!\eta , \forall u \!\in\! \mathcal{U},
\\
&     \xi_{k,u}\in \left\{ 0,1 \right\} , \forall u\in \mathcal{U},\forall k\in \mathcal{K},
\\
&    \sum_{u\in \mathcal{U}}{\xi_{k,u}}=1, \forall k\in \mathcal{K},
\\
&    \text{(12b), (12c)}, \nonumber
\end{align}
\end{subequations}
where (13b) guarantees minimum sensing SINR, and (13c) and (13d) enforce binary beam pattern assignments with one beam pattern per target. (P2) is a mixed-integer non-convex problem due to unit-modulus constraints, binary variables, fractional SINR expressions, and closely coupled optimization variables, which cannot be solved directly using existing methods.

\section{RALM-Based Optimization Algorithm for (P2) }
In this section, we propose an efficient Riemannian manifold optimization algorithm for (P2), recasting the SINR constraints and binary scheduling variables to fit a smooth manifold-constrained formulation.

\vspace{-4pt}
\subsection{Product Manifold Constructing}
\vspace{-2pt}
The unit-modulus constraints $|\phi_m| = 1, \forall m \in \mathcal{M}$ and $|\theta_n| = 1, \forall n \in \mathcal{N}$ in $\boldsymbol{\phi}$ and $\boldsymbol{\theta}$ inherently define a Riemannian manifold known as the \textit{complex circle manifold}. However, the beam pattern scheduling variables $\xi_{k,u}$ are originally binary and do not directly constitute a manifold. To tackle this difficulty, we first relax the discrete binary variables $\{\xi_{k,u}\}$ into continuous variables within the unit interval
\begin{align}
\xi_{k,u}\in \left\{ 0,1 \right\} \Rightarrow 1\ge \xi_{k,u}\ge 0, \,  \forall k \in \mathcal{K}, \forall u \in \mathcal{U}.
\end{align}
Taking into account the additional constraint $\sum_{u=1}^U{\xi_{k,u}}=1,\forall k \in \mathcal{K}$ in (13d) and enforcing strict positivity, the relaxed variables, denoted by $\boldsymbol{\varXi}=[\xi_{k,u}]_{K\times U}$, reside within a probability simplex, forming a \emph{multinomial manifold} as [29]
\begin{align}
\sum_{u=1}^U{\xi_{k,u}}=1, \xi_{k,u} > 0, \,  \forall k \in \mathcal{K}, \forall u \in \mathcal{U}.
\end{align}
In general, relaxing the binary beam pattern indicators $\xi_{k,u}$ to the simplex probability simplex formally allows fractional results. However, the structure of effective SINR expressions inherently pushes every row of $\boldsymbol{\varXi}$ towards a one-hot vector in the optimization process. Because, for each user $k$, the optimization favors $\xi_{k,u}$ being one for the beam patterns $u$ that yield the largest $\gamma_{k,u}$ and drives all the others to near zero. Consequently, the relaxed variables converge to quasi-binary values during the manifold optimization iterations, with non-selected entries typically falling below a very small, e.g., $10^{-6}$, order according to the predetermined numerical tolerance. A final rounding step, therefore, restores strict binary scheduling with virtually no performance loss.

Accordingly, in addition to the Euclidean space $\mathcal{R}_{\eta}: \eta \in \mathbb{R}$, we define feasible sets of optimization variables as the following manifolds 
\begin{subequations}
\begin{align}
    &\mathcal{R}_{\boldsymbol{\phi}} = \left\{ \boldsymbol{\phi} \in \mathbb{C}^M : |\phi_m| = 1, \forall m \in \mathcal{M} \right\}, 
    \\
    &\mathcal{R}_{\boldsymbol{\theta}} = \left\{ \boldsymbol{\theta} \in \mathbb{C}^N : |\theta_n| = 1, \forall n \in \mathcal{N}\right\}, 
    \\
    &\mathcal{R}_{\boldsymbol{\varXi}} \! =\! \Big\{ \!\boldsymbol{\varXi} \in \mathbb{R}^{K \times U}\!:\! \sum_{u\in \mathcal{U}} \xi_{k,u}\! =\!1, \xi_{k,u} \!>\!0, \forall u \in \mathcal{U}, \forall k \in \mathcal{K}  \Big\}.
\end{align}
\end{subequations}
A manifold can be interpreted as a topological space that locally resembles a Euclidean space, and a tangent vector describes the direction in which a point can be updated on the manifold. All tangent vectors at a given point, representing all possible directions in which the point can move, collectively form the tangent space, which can be explicitly defined as
\begin{subequations}
\begin{align}
&\mathcal{T}_{\boldsymbol{\phi }}=\big\{\boldsymbol{t}\in \mathbb{C}^M: \Re\left\{ \phi _m^* t_m \right\} =0,\forall m\in \mathcal{M}\big\},
\\
&\mathcal{T}_{\boldsymbol{\theta }}=\big\{\boldsymbol{t}\in \mathbb{C}^N: \Re\left\{ \theta _n^* t_n \right\} =0,\forall n\in \mathcal{N}\big\},
\\
&\mathcal{T}_{\boldsymbol{\varXi}}=\Big\{ \boldsymbol{T}\in \mathbb{R}^{K\times U}: \sum_{u \in \mathcal{U}}{t_{k,u}}=0,\forall u\in \mathcal{U}, \forall k\in \mathcal{K} \Big\},\!\!
\end{align}
\end{subequations}
where condition $\sum_{u \in \mathcal{U}}{t_{k,u}}=0$ ensures that any infinitesimal perturbation within the tangent space does not violate the simplex constraint that the components sum to one. In addition, the tangent space $\mathcal{T}_{\eta}$ of the Euclidean space $\mathcal{R}_{\eta}$ is itself. Then, by taking the Cartesian product of the individual manifolds for optimization variables, we construct a product manifold defined as 
\begin{align}
    \mathcal{R}_{\boldsymbol{z}} = \mathcal{R}_{\eta} \times\mathcal{R}_{\boldsymbol{\phi}} \times \mathcal{R}_{\boldsymbol{\theta}} \times \mathcal{R}_{\boldsymbol{\varXi}},
\end{align}
where $\times$ denotes the Cartesian product between sets and $\boldsymbol{z}:=\left( \boldsymbol{\phi},\boldsymbol{\theta},\boldsymbol{\varXi},\eta \right)$ denotes the optimization variables. For any point $\boldsymbol{z} \in \mathcal{R}_{\boldsymbol{z}}$, the tangent space at that point is the direct sum of the individual tangent spaces
\begin{align}
\mathcal{T}_{\boldsymbol{z}} = \mathcal{T}_{\eta} \oplus\mathcal{T}_{\boldsymbol{\phi}} \oplus \mathcal{T}_{\boldsymbol{\theta}} \oplus \mathcal{T}_{\boldsymbol{\varXi}},
\end{align}
where $\oplus$ denotes the direct sum of vector spaces.

Then, we can equivalently recast (P2) as 
\begin{subequations}
\begin{align}
\text{(P2.1)} :&\underset{ \boldsymbol{z} \in \mathcal{R}_{\boldsymbol{z}}}{\max}\,\,\eta 
\\
\text{s.t.} \,\,& \eta -\sum_{u\in \mathcal{U}}{\xi_{k,u}\frac{\beta _{k}^{2}\left( \boldsymbol{v}_{u}^{H}\boldsymbol{G}_k\boldsymbol{v}_u \right) ^2}{\sum_{i\in \mathcal{K},i\ne k}{\beta _{i}^{2}\left( \boldsymbol{v}_{u}^{H}\boldsymbol{G}_i\boldsymbol{v}_u \right) ^2}+\frac{\sigma _{k}^{2}}{PL^2}}}\le 0,\forall u,
\end{align}
\end{subequations}
which is a constrained manifold optimization problem.

\subsection{RALM Framework for Constrained Manifold Optimization}

\subsubsection{Fundamentals}
RALM generalizes the classical Powell–Hestenes–Rockafellar framework of augmented Lagrange methods to problems whose optimization variables lie on a smooth manifold $\mathcal{R}_{\boldsymbol{z}}$ [30]. Consider problems of the form
\begin{subequations}
\begin{align}
&\underset{\boldsymbol{z}\in \mathcal{R}_{\boldsymbol{z}}}{\min}\,\,f\left( \boldsymbol{z} \right) \,\,
\\
\text{s.t.} \,\, & h_i\left( \boldsymbol{z} \right) =0, i\in \mathcal{I},
\\
& g_j\left( \boldsymbol{z} \right) \le 0, j\in \mathcal{J},
\end{align}
\end{subequations}
where $f\left( \boldsymbol{z} \right)$ and all constraint functions are continuously differentiable. RALM solves it by alternating between an inner step and an outer step. Specifically, the inner step solves an unconstrained, manifold-restricted subproblem on $\mathcal{R}_{\boldsymbol{z}}$ with an augmented Lagrangian, while the outer step updates Lagrange multipliers and the penalty coefficient that penalizes constraint violations. With equality multipliers $\boldsymbol{\mu} \in \mathbb{R}_{+}^{\left| \mathcal{I} \right|}$, inequality multipliers $\boldsymbol{\lambda} \in \mathbb{R}_{+}^{\left| \mathcal{J} \right|}$, and a penalty coefficient $\rho >0$, the manifold-restricted augmented Lagrangian reads
\begin{align}
\mathcal{L}_{\rho}\left( \boldsymbol{\varXi},\boldsymbol{\lambda },\boldsymbol{\mu } \right) =&f\left( \boldsymbol{z} \right)  +\frac{\rho}{2}\Bigg( \sum_{i\in \mathcal{I}}{\left( \frac{\mu _i}{\rho}+h_i\left( \boldsymbol{z} \right) \right)}^2 \nonumber
\\
&+\sum_{j\in \mathcal{J}}{\max \left\{ 0,\frac{\lambda _j}{\rho}+g_j\left( \boldsymbol{z} \right) \right\} ^2} \Bigg), 
\end{align}
which is identical to its counterpart of the Euclidean formula, except that $\boldsymbol{z}$ is constrained to remain on $\mathcal{R}_{\boldsymbol{z}}$. Hence, each inner minimization can exploit the full toolbox of Riemannian optimization, including retractions, vector transports, and Riemannian gradients or Hessians. 
Specifically, at any iteration, the Riemannian gradient w.r.t. $\boldsymbol{z}$ is obtained by projecting the Euclidean gradient of $\mathcal{L}_{\rho}$ onto the tangent space $\mathcal{T}_{\boldsymbol{z}}$
\begin{align}
\nabla_{\mathcal{R}_{\boldsymbol{z}}}\mathcal{L}_{\rho}(\boldsymbol{z})
    = \mathsf{Proj}_{\mathcal{T}_{\boldsymbol{z}}}
      \left[\nabla_{\boldsymbol{z}}\mathcal{L}_{\rho}(\boldsymbol{z})\right].
\end{align}
The outer loop then increases~$\rho$ and updates $(\boldsymbol{\lambda},\boldsymbol{\mu})$ until both primal feasibility and dual convergence criteria are met. The overall algorithm operated in a nested-iterative manner. Thanks to the manifold projection, no extra feasibility restoration is required after each inner iteration, and the algorithm inherits the convergence guarantees of its Euclidean analogue under standard regularity conditions [30].

\subsubsection{Minimizing augmented Lagrangian in inner loop}

With multipliers $\boldsymbol{\lambda}$, the augmented Lagrangian of (P2.1) can be written as
\begin{align}
\!\!\mathcal{L}_{\rho}\left( \boldsymbol{z},\boldsymbol{\lambda } \right) =-\eta +\frac{\rho}{2}\sum_{k\in \mathcal{K}}{\max \left\{ 0,\frac{\lambda _k}{\rho}+q_k\left( \boldsymbol{\phi },\boldsymbol{\theta },\boldsymbol{\varXi } \right) \right\} ^2}\!, \!\!
\end{align}
where
\begin{subequations}
\begin{align}
&q_k\left( \boldsymbol{z} \right) =\eta -\sum_{u\in \mathcal{U}}{\xi_{k,u}\gamma _{k,u}}\left( \boldsymbol{\phi },\boldsymbol{\theta },\boldsymbol{\varXi } \right), 
\\
&\gamma _{k,u}\left( \boldsymbol{\phi },\boldsymbol{\theta },\boldsymbol{\varXi } \right) =\frac{\beta _{k}^{2}a_{k,u}^{2}\left( \boldsymbol{\phi },\boldsymbol{\theta },\boldsymbol{\varXi } \right)}{b_{k,u}\left( \boldsymbol{\phi },\boldsymbol{\theta },\boldsymbol{\varXi } \right)}
\\
&b_{k,u}\left( \boldsymbol{\phi },\boldsymbol{\theta },\boldsymbol{\varXi } \right)=\sum_{i\in \mathcal{K},i\ne k}{\beta _{i}^{2}a_{i,u}^{2}\left( \boldsymbol{\phi },\boldsymbol{\theta },\boldsymbol{\varXi } \right)}+\frac{\sigma _{k}^{2}}{PL^2}
\\
&a_{k,u}\left( \boldsymbol{\phi },\boldsymbol{\theta },\boldsymbol{\varXi } \right) =\boldsymbol{v}_{u}^{H}\boldsymbol{G}_k\boldsymbol{v}_u
\end{align}
\end{subequations}
With multipliers $\boldsymbol{\lambda}$ and penalty coefficient $\rho$ fixed, (P2.1) is converted into a standard unconstrained manifold optimization problem as follows
\begin{align}
\text{(P2.2)} : \underset{\boldsymbol{z}\in \mathcal{R}_{\boldsymbol{z}}}{\min}\,\,\mathcal{L}_{\rho}\left( \boldsymbol{z} \right), 
\end{align}
which can be solved via an efficient Riemannian conjugate-gradient (RCG) algorithm, detailed later in the next subsection.

\subsubsection{Updating multipliers and penalty in outer loop} 
After obtaining an approximate minimizer $\boldsymbol{z}^{(\ell+1)}$ of (P2.2), the corresponding Euclidean augmented Lagrangian method is easily extended to the Riemannian case as well. Here, we define the clip operator as
\begin{align}
\mathsf{Clip}_{\left[ \mathfrak{a},\mathfrak{b} \right]}\left( \mathfrak{c} \right) =\max \left\{ \mathfrak{a},\min \left\{ \mathfrak{b},\mathfrak{c} \right\} \right\} 
\end{align}
Then, the multipliers are updated via a safeguarded clipping rule
\begin{align}
\lambda _{k}^{\left( \ell +1 \right)}=\mathsf{Clip}_{\left[ \lambda _{k}^{\min},\lambda _{k}^{\max} \right]}\left( \lambda _{k}^{\left( \ell \right)}+\rho ^{\left( \ell \right)}q_k\left( \boldsymbol{z}^{\left( \ell +1 \right)} \right) \right) ,
\end{align}
where $\lambda _{k}^{\min}\in \mathbb{R}^K $ and $\lambda _{k}^{\max} \in \mathbb{R}^K$ with $\lambda _{k}^{\min}\leq \lambda _{k}^{\max}$ represent multiplier boundaries. The penalty coefficient is increased only when constraint violations shrink fast enough, helping alleviate the effect of ill conditioning and improving robustness [], following the conditioned update rule
\begin{align}
\rho ^{\left( \ell +1 \right)}=\begin{cases}
	\varsigma _{\rho}\rho ^{\left( \ell \right)}, \text{if} \max _{k\in \mathcal{K}}\left\{ \iota _{k}^{\left( \ell +1 \right)} \right\} \le \varsigma _{\iota}\max _{k\in \mathcal{K}}\left\{ \iota _{k}^{\left( \ell \right)} \right\}\\
	\rho ^{\left( \ell \right)}, \text{otherwise}.\\
\end{cases}
\end{align}
with constant $\varsigma _{\rho}\ge 1$, ratio $\varsigma _{\iota}\in (0,1)$, and auxiliary parameter 
\begin{align}
\iota _{k}^{\left( \ell +1 \right)}=\max \left\{ q_k\left( \boldsymbol{z}^{\left( \ell +1 \right)} \right) ,-\frac{\lambda _{k}^{\left( \ell \right)}}{\rho ^{\left( \ell \right)}} \right\}.
\end{align}

\subsection{RCG Method for (P2.2)}
To solve (P2.2), we adopt the RCG method, which generalizes the classical conjugate gradient algorithm to optimization over curved spaces, that is, the product manifold defined in (33). The RCG method proceeds by computing the Riemannian gradient, determining the conjugate search direction, and applying a retraction operation to update variables and ensure that updated variables remain on the manifold.

\subsubsection{Computing the Riemannian gradient}
Before determining the search direction, we need to calculate the Riemannian gradient, a vector field on the manifold $ \mathcal{R}$ obtained by projecting the Euclidean gradient onto the tangent space of the manifold. The explicit Euclidean gradients of $\mathcal{L}_{\rho}\left( \boldsymbol{z} \right)$ w.r.t. variables $\eta$, $\boldsymbol{\phi}$, $\boldsymbol{\theta}$, and $\boldsymbol{\varXi}$ are given, respectively, as follows. If $\frac{\lambda _k}{\rho}+q_k\le 0$, we have $\mathcal{L}_{\rho}\left( \boldsymbol{z} \right) =-\eta $, and thereby 
\begin{align}
\nabla_{\eta}\mathcal{L}=-1, \, \frac{\partial \mathcal{L}}{\partial \xi_{k,u}}=0, \, \nabla _{\boldsymbol{\phi }}\mathcal{L}=0, \,\nabla _{\boldsymbol{\theta }}\mathcal{L}=0,
\end{align}
Otherwise, if $\frac{\lambda _k}{\rho}+q_k>0$, we have
\begin{align}
\mathcal{L}_{\rho}\left( \boldsymbol{z}\right) =-\eta +\sum_{k\in \mathcal{K}}{\lambda _kq_k}+\sum_{k\in \mathcal{K}}{\frac{\rho}{2}q_{k}^{2}}+\sum_{k\in \mathcal{K}}{\frac{\lambda _{k}^{2}}{2\rho}},
\end{align}
and the corresponding Euclidean gradients are
\begin{subequations}
\begin{align}
&\nabla _{\eta}\mathcal{L}=-1+\sum_{k\in \mathcal{K}}{\chi _k}
\\
&\nabla_{\boldsymbol{\varXi}} \mathcal{L}=-\left[\chi _k\gamma _{k,u}\right]_{K \times U}
\\
&\nabla _{\boldsymbol{\phi }}\mathcal{L}=-\sum_{k\in \mathcal{K}}{\chi _k\sum_{u\in \mathcal{U}}{\xi_{k,u}\nabla _{\boldsymbol{\phi }}\gamma _{k,u}}}
\\
&\nabla _{\boldsymbol{\theta }}\mathcal{L}=-\sum_{k\in \mathcal{K}}{\chi _k\sum_{u\in \mathcal{U}}{\xi_{k,u}\nabla _{\boldsymbol{\theta }}\gamma _{k,u}}}
\end{align}
\end{subequations}
where $\chi _k=\lambda _k+\rho q_k$ denotes the effective coefficients and the function $\nabla \mathcal{L}_{\rho}(\boldsymbol{z})$ is abbreviated as $\nabla\mathcal{L}$ for brevity, and the related partial differentials are computed as
\begin{subequations}
\begin{align}
&\nabla _{\boldsymbol{\phi }}\gamma _{k,u}=\frac{\beta _{k}^{2}}{b_{k,u}^{2}}\left( b_{k,u}^{}\nabla _{\boldsymbol{\phi }}a_{k,u}^{2}-a_{k,u}^{2}\nabla _{\boldsymbol{\phi }}b_{k,u}^{} \right) 
\\
&\nabla _{\boldsymbol{\theta }}\gamma _{k,u}\!=\!\frac{2\beta _{k}^{2}}{b_{k,u}^{2}}\Big( \!a_{k,u}b_{k,u}\!\nabla _{\boldsymbol{\theta }}a_{k,u}\!-\!a_{k,u}^{2}\!\sum_{k'\in \mathcal{K}\backslash k}{\!\beta _{k'}^{2}a_{k',u}\!\nabla _{\boldsymbol{\theta }}a_{k',u}^{}}\! \Big) 
\\
&\nabla _{\boldsymbol{\phi }}a_{k,u}^{}=2\mathrm{diag}\left( \boldsymbol{S}_u\boldsymbol{\theta }+\boldsymbol{e}_u \right) ^H\boldsymbol{G}_k\boldsymbol{v}_u
\\
&\nabla _{\boldsymbol{\theta }}a_{k,u}^{}=2\boldsymbol{S}_{u}^{H}\mathrm{diag}\left( \boldsymbol{\phi } \right) ^H\boldsymbol{G}_k\boldsymbol{v}_u
\\
&\nabla _{\boldsymbol{\phi }}b_{k,u}^{}=2\sum_{k'\in \mathcal{K}\backslash k}{\beta _{k'}^{2}a_{k',u}^{}\nabla _{\boldsymbol{\phi }}a_{k',u}^{}}
\\
&\nabla _{\boldsymbol{\theta }}b_{k,u}^{}=2\sum_{k'\in \mathcal{K}\backslash k}{\beta _{k'}^{2}a_{k',u}^{}\nabla _{\boldsymbol{\theta }}a_{k',u}^{}}
\end{align}
\end{subequations}

Having computed the Euclidean gradients, we project them onto the tangent spaces of their respective manifolds to obtain the Riemannian gradients. Specifically, the Riemannian gradient $\nabla_{\mathcal{R}_{\boldsymbol{\phi}}} \mathcal{L}$ is obtained by projecting the Euclidean gradient $\nabla_{\boldsymbol{\phi}} \mathcal{L}$ onto the tangent space $\mathcal{T}_{\boldsymbol{\phi}}$
\begin{subequations}
\begin{align}
\nabla_{\mathcal{R}_{\boldsymbol{\phi}} } \mathcal{L} &= \mathsf{Proj}_{\mathcal{T}_{\boldsymbol{\phi}}}(\nabla_{\boldsymbol{\phi}} \mathcal{L}) 
\\
&= \nabla_{\boldsymbol{\phi}} \mathcal{L} - \Re\left( \nabla_{\boldsymbol{\phi}} \mathcal{L} \odot \boldsymbol{\phi}^* \right) \odot \boldsymbol{\phi},
\end{align}
\end{subequations}
where $\mathsf{Proj}_{\boldsymbol{\phi}}(\cdot)$ denotes the projection operation, $\odot$ denotes the Hadamard product, and $\boldsymbol{\phi}^*$ is the complex conjugate of $\boldsymbol{\phi}$.
Similarly, the Riemannian gradient $\nabla_{\mathcal{R}_{\boldsymbol{\theta}}} \mathcal{L} $ is given by
\begin{align}
\nabla_{\mathcal{R}_{\boldsymbol{\theta}} }\mathcal{L} = \nabla_{\boldsymbol{\theta}} \mathcal{L}  - \Re\left( \nabla_{\boldsymbol{\theta}} \mathcal{L}  \odot \boldsymbol{\theta}^* \right) \odot \boldsymbol{\theta}.
\end{align}
For the beam pattern scheduling matrix $\boldsymbol{\varXi}$ in the multinomial manifold, the Riemannian gradient $\nabla_{\mathcal{R}_{\boldsymbol{\varXi}}} \mathcal{L} $ is obtained by 
\begin{subequations}
\begin{align}
\nabla_{\mathcal{R}_{\boldsymbol{\varXi}} } \mathcal{L}  
&= \mathsf{Proj}_{\mathcal{T}_{\boldsymbol{\varXi}}}(\nabla_{\boldsymbol{\varXi}} \mathcal{L} ) 
\\
&= \nabla_{\boldsymbol{\varXi}} \mathcal{L}  - \Big( \frac{1}{U}\nabla_{\boldsymbol{\varXi}} \mathcal{L} \cdot \boldsymbol{1}_{U\times1} \Big) \cdot \boldsymbol{1}_{U\times1}^\mathrm{T},
\end{align}
\end{subequations}
where $\boldsymbol{1}_{U\times1}$ is a column vector of length $U$, and the subtraction ensures that each row of $\nabla_{\mathcal{R}_{\boldsymbol{\varXi}} } \mathcal{L}$ sums to zero, satisfying the tangent space conditions in (17c). In addition, we have $\nabla_{\mathcal{R}_{\eta}}\mathcal{L}= \nabla_{\eta}\mathcal{L}$ because many operations such as projection become identity mappings on the Euclidean manifold.

\subsubsection{Determining Conjugate Descent Direction}
To determine the descent direction on the manifold, we employ the conjugate gradient method adapted to Riemannian manifolds. Specifically, in iteration $i$, the descent directions $d_{\eta}^{(i)}$, $\boldsymbol{d}_{\boldsymbol{\phi}}^{(i)}$, $\boldsymbol{d}_{\boldsymbol{\theta}}^{(i)}$, and $\boldsymbol{d}_{\boldsymbol{\varXi}}^{(i)}$ are calculated using the Polak-Ribiere formula. Specifically, taking $\boldsymbol{\phi}$ as an example, let $\nabla _{\mathcal{R}_{\boldsymbol{\phi }}}\mathcal{L}^{(i)}$ denote its Riemannian gradient in iteration $i$. The conjugate gradient update coefficient $\varrho_{\boldsymbol{\phi}}^{(i)}$, which adjusts the new search direction accounting for the curvature of the manifold, is computed as
\begin{subequations}
\begin{equation}
\varrho _{\boldsymbol{\phi }}^{\left( i \right)}=\frac{\left< \nabla _{\mathcal{R}_{\boldsymbol{\phi }}}\mathcal{L}^{\left( i \right)},\nabla _{\mathcal{R}_{\boldsymbol{\phi }}}\mathcal{L}^{\left( i \right)}-\nabla _{\mathcal{R}_{\boldsymbol{\phi }}}\mathcal{L}^{\left( i-1 \right)} \right>}{ \left< \nabla _{\mathcal{R}_{\boldsymbol{\phi }}}\mathcal{L}^{\left( i-1 \right)},\nabla _{\mathcal{R}_{\boldsymbol{\phi }}}\mathcal{L}^{\left( i-1 \right)}  \right>},
\end{equation}
where $\left\langle \cdot, \cdot \right\rangle$ denotes the Riemannian metric, which is Euclidean inner product in this case. The descent direction is then updated as
\begin{align}
\boldsymbol{d}_{\boldsymbol{\phi}}^{(i)} = &-\nabla _{\mathcal{R}_{\boldsymbol{\phi }}}\mathcal{L}^{\left( i \right)}+ \varrho_{\boldsymbol{\phi}}^{(i)} \mathsf{Tran}_{\mathcal{T}_{\boldsymbol{\phi }}^{\left( i \right)}}\left( \boldsymbol{d}_{\boldsymbol{\phi }}^{\left( i-1 \right)} \right) ,
\\
=&-\nabla _{\mathcal{R}_{\boldsymbol{\phi }}}\mathcal{L}^{\left( i \right)}+ \varrho_{\boldsymbol{\phi}}^{(i)} \mathsf{Proj}_{\mathcal{T}_{\boldsymbol{\phi }}^{\left( i \right)}}\left( \boldsymbol{d}_{\boldsymbol{\phi }}^{\left( i-1 \right)} \right) ,
\end{align}
\end{subequations}
where $\mathsf{Tran}_{\mathcal{T}_{\boldsymbol{\phi }}^{\left( i \right)}}\left( \boldsymbol{d}_{\boldsymbol{\phi }}^{\left( i-1 \right)} \right) $ is the vector transport operation that moves the previous search direction $\boldsymbol{\eta}^{(i-1)}$, also a tangent vector, from the tangent space at $\boldsymbol{\phi}^{(i-1)}$ to $\boldsymbol{\phi}^{(i)}$, ensuring appropriate mapping with the current tangent space and accounting for the curvature of the manifold. For complex circle manifolds, the transport operation is exactly the same as the projection operation. Similarly, the update coefficient and descent direction for $\boldsymbol{\theta}$ are
\begin{subequations}
\begin{align}
&\varrho _{\boldsymbol{\theta }}^{\left( i \right)}=\frac{\left< \nabla _{\mathcal{R}_{\boldsymbol{\theta }}}\mathcal{L}^{\left( i \right)},\nabla _{\mathcal{R}_{\boldsymbol{\theta }}}\mathcal{L}^{\left( i \right)}-\nabla _{\mathcal{R}_{\boldsymbol{\theta }}}\mathcal{L}^{\left( i-1 \right)} \right>}{\left< \nabla _{\mathcal{R}_{\boldsymbol{\theta }}}\mathcal{L}^{\left( i-1 \right)},\nabla _{\mathcal{R}_{\boldsymbol{\theta }}}\mathcal{L}^{\left( i-1 \right)} \right>},
\\
&\boldsymbol{d}_{\boldsymbol{\theta}}^{(i)} = -\nabla _{\mathcal{R}_{\boldsymbol{\theta }}}\mathcal{L}^{\left( i \right)}+ \varrho_{\boldsymbol{\theta}}^{(i)} \mathsf{Proj}_{\mathcal{T}_{\boldsymbol{\theta }}^{\left( i \right)}}\left( \boldsymbol{d}_{\boldsymbol{\theta }}^{\left( i-1 \right)} \right) ,.
\end{align}
\end{subequations}
For the matrix variable $\boldsymbol{\varXi}$ in the multinomial manifold, the conjugate gradient update coefficient $\varrho_{\boldsymbol{\varXi}}^{(i)}$ is computed as
\begin{subequations}
\begin{equation}
\varrho _{\boldsymbol{\varXi}}^{\left( i \right)}=\frac{\left< \nabla _{\mathcal{R}_{\boldsymbol{\varXi}}}\mathcal{L}^{\left( i \right)},\nabla _{\mathcal{R}_{\boldsymbol{\varXi}}}\mathcal{L}^{\left( i \right)}-\nabla _{\mathcal{R}_{\boldsymbol{\varXi}}}\mathcal{L}^{\left( i-1 \right)} \right>_\mathrm{F}}{\left< \nabla _{\mathcal{R}_{\boldsymbol{\varXi}}}\mathcal{L}^{\left( i-1 \right)},\nabla _{\mathcal{R}_{\boldsymbol{\varXi}}}\mathcal{L}^{\left( i-1 \right)} \right>_\mathrm{F}},
\end{equation}
where the Riemannian metric $\left\langle \cdot, \cdot \right\rangle_\mathrm{F}$ here is the Frobenius inner product that multiplies the entries of two matrices and sums them up. The descent direction is then given by
\begin{align}
\boldsymbol{d}_{\boldsymbol{\varXi}}^{(i)} = &
-\nabla _{\mathcal{R}_{\boldsymbol{\varXi}}}\mathcal{L}^{\left( i \right)}+\varrho_{\boldsymbol{\varXi}}^{(i)} \mathsf{Tran}_{\mathcal{T}_{\boldsymbol{\varXi}}^{\left( i \right)}}\left( \boldsymbol{d}_{\boldsymbol{\varXi}}^{\left( i-1 \right)} \right) 
\\
&=-\nabla _{\mathcal{R}_{\boldsymbol{\varXi}}}\mathcal{L}^{\left( i \right)}+ \varrho_{\boldsymbol{\varXi}}^{(i)} \boldsymbol{d}_{\boldsymbol{\varXi}}^{(i-1)},
\end{align}
\end{subequations}
where the vector transport of the previous search direction between tangent spaces reduces to the identity mapping since the multinomial manifold has a flat geometric structure. Finally, the conjugate descent direction for $\eta^{(i)}$ completely reduces to the case in Euclidean space with identity vector transport operation, written as follow
\begin{subequations}
\begin{align}
&\varrho _{\eta}^{\left( i \right)}=\frac{\nabla _{\eta}\mathcal{L}^{\left( i \right)}\left( \nabla _{\eta}\mathcal{L}^{\left( i \right)}-\nabla _{\eta}\mathcal{L}^{\left( i-1 \right)} \right)}{\left( \nabla _{\eta}\mathcal{L}^{\left( i-1 \right)} \right) ^2},
\\
&d_{\eta}^{\left( i \right)}=-\nabla _{\eta}\mathcal{L}^{\left( i \right)}+\varrho _{\eta}^{\left( i \right)}d_{\eta}^{\left( i-1 \right)}.
\end{align}
\end{subequations}

\subsubsection{Retraction Operation}
After updating the variables by moving along these descent directions, we map the new points back onto the manifold using retraction operations $\mathsf{Retr}(\cdot)$. This retraction ensures that the updated variables satisfy their manifold constraints, allowing the optimization to proceed within the feasible region. Specifically,
The updated variable $\boldsymbol{\phi}^{(i+1)}$ is computed as
\begin{subequations}
\begin{align}
\boldsymbol{\phi }^{\left( i+1 \right)}&=\mathsf{Retr}_{\boldsymbol{\phi }^{\left( i \right)}}( \alpha_{\boldsymbol{\phi}} ^{\left( i \right)}\boldsymbol{d}_{\boldsymbol{\phi}}^{(i)} ) 
\\
&=\left[ \frac{( \boldsymbol{\phi }^{\left( i \right)}+\alpha_{\boldsymbol{\phi}}^{\left( i \right)}\boldsymbol{d}_{\boldsymbol{\phi}}^{(i)}) _m}{| ( \boldsymbol{\phi }^{\left( i \right)}+\alpha_{\boldsymbol{\phi}}^{\left( i \right)}\boldsymbol{d}_{\boldsymbol{\phi}}^{(i)} ) _m |} \right] ,
\end{align}
\end{subequations}
where $\alpha_{\boldsymbol{\phi}}^{(i)}$ is the step size determined by a line search method, and the element-wise normalization ensures that $\boldsymbol{\phi}^{(i+1)}$ lies on the complex circle manifold.
Similarly, the update for $\boldsymbol{\theta}$ is
\begin{align}
\boldsymbol{\theta }^{\left( i+1 \right)}=\left[ \frac{( \boldsymbol{\theta }^{\left( i \right)}+\alpha_{\boldsymbol{\theta}}^{\left( i \right)}\boldsymbol{d}_{\boldsymbol{\theta}}^{(i)})_n}{| ( \boldsymbol{\theta }^{\left( i \right)}+\alpha_{\boldsymbol{\theta}}^{\left( i \right)}\boldsymbol{d}_{\boldsymbol{\theta}}^{(i)} )_n |} \right].
\end{align}
For $\boldsymbol{\varXi}$, the retraction involves projecting the updated $\boldsymbol{\varXi}$ back onto the multinomial manifold
\begin{subequations}
\begin{align}
\boldsymbol{\varXi}^{(i+1)}&=\mathsf{Retr}_{\boldsymbol{\varXi}^{\left( i \right)}}( \alpha_{\boldsymbol{\varXi}}^{\left( i \right)}\boldsymbol{d}_{\boldsymbol{\varXi}}^{(i)} ) 
\\
& = \Pi_{\mathcal{R}_{\boldsymbol{\varXi}}}\big( \boldsymbol{\varXi}^{(i)} + \alpha_{\boldsymbol{\varXi}}^{\left( i \right)} \boldsymbol{d}_{\boldsymbol{\varXi}}^{(i)} \big),
\end{align}
\end{subequations}
where $\Pi_{\mathcal{R}_{\boldsymbol{\varXi}}}$ denotes the projection onto the multinomial manifold, which can be performed using the algorithm in [35]. For $\eta$, we still have the identity mapping
\begin{align}
\eta ^{\left( i+1 \right)}=\eta ^{\left( i \right)}+\alpha _{\eta}^{\left( i \right)}d_{\eta}^{\left( i \right)}.
\end{align}

\renewcommand{\algorithmicrequire}{\textbf{Input:}}
\renewcommand{\algorithmicensure}{\textbf{Output:}}
\begin{algorithm}[!t]
\caption{Inner RCG for solving the subproblem (P2.2)}
\label{alg:general_solution}
\begin{algorithmic}[1]
\REQUIRE $N_c, N_r, M_c, M_r, K, U,\{\boldsymbol{S}_u\}, \{\boldsymbol{e}_u\}$, and $\epsilon_{\text{RCG}}$
\ENSURE $\boldsymbol{z}^\star: (\eta^\star, \boldsymbol{\phi}^\star, \boldsymbol{\theta}^\star,\boldsymbol{\varXi}^\star)$
\STATE Initialize $\eta^{(0)}$, $\boldsymbol{\phi}^{(0)}$, $\boldsymbol{\theta}^{(0)}$, and $\boldsymbol{\varXi}^{(0)}$ within their respective manifolds; Set initial descent directions $\boldsymbol{d}_{\eta}^{(0)}$, $\boldsymbol{d}_{\boldsymbol{\phi}}^{(0)}$, $\boldsymbol{d}_{\boldsymbol{\theta}}^{(0)}$, and $\boldsymbol{d}_{\boldsymbol{\varXi}}^{(0)}$; Set iteration index $i = 0$.
        \WHILE{$i=0$ \OR $\|\nabla_{\mathcal{R}_{\boldsymbol{z}}}\mathcal{L}_{\rho}(\boldsymbol{z}^{(i)}) \| \geq \epsilon_{\text{RCG}}$}
            \STATE Calculate step size $\alpha_{\eta}^{(i)}$, $\alpha_{\boldsymbol{\phi}}^{(i)}$, $\alpha_{\boldsymbol{\theta}}^{(i)}$, and $\alpha_{\boldsymbol{\varXi}}^{(i)}$ using backtracking algorithms [34].
            \STATE Update $\boldsymbol{\phi}^{(i+1)}$ using (33c), (35), (38), and (42).
            \STATE Update $\boldsymbol{\theta}^{(i+1)}$ using (33d), (36), (39), and (43).
            \STATE Update $\boldsymbol{\varXi}^{(i+1)}$ using (33b), (37), (40), and (44).
            \STATE Update $\eta^{(i+1)}$ using (33a), (41), and (45).
            \STATE $i \leftarrow  i + 1$
        \ENDWHILE
\RETURN $\eta^\star=\eta^{(i)}, \boldsymbol{\phi}^\star=\boldsymbol{\phi}^{(i)}$, $\boldsymbol{\theta}^\star=\boldsymbol{\theta}^{(i)}$, and $\boldsymbol{\varXi}^\star=\boldsymbol{\varXi}^{(i)}$.
\end{algorithmic}
\end{algorithm}

\vspace{-12pt}
\subsection{Overall Algorithm, Complexity, and Convergence}
\textbf{Algorithm 1} summarizes the RCG method to solve (P2.2), initialized with feasible points within each variable's respective manifolds, and iteratively updating all variables until convergence. RCG guarantees convergence to a stationary point under standard regularity conditions with Lipschitz continuity of gradients and appropriate line search steps [34]. At convergence, the manifold structures preserve unit-modulus (12b) and (12c) and the probability simplex constraints (15), and thresholding techniques are applied to the scheduling matrix to recover binary decisions. Per iteration, the dominant computation is to evaluate quadratic forms $a_{k,u} = \boldsymbol{v}_u^H \boldsymbol{G}_k \boldsymbol{v}_u$ for all user–pattern pairs $(k, u)$, which costs $\mathcal{O}(KUM^2)$ for dense $\boldsymbol{G}_k \in \mathbb{C}^{M\times M}$. Then, assembling the Euclidean gradients $\nabla _{\boldsymbol{\theta}}$ and $\nabla _{\boldsymbol{\phi}}$ through the chain rules provided, as well as $\nabla _{\eta}$ and $\nabla _{\boldsymbol{\varXi}}$, incurs $\mathcal{O}(KU(M+N))$. Projecting onto tangent spaces to obtain Riemannian gradients, Polak–Ribière updates and vector transports require $\mathcal{O}(KU+M+N)$ for each. Retractions for $\boldsymbol{\theta}$ and $\boldsymbol{\phi}$ are element-wise normalizations $\mathcal{O}(M+N)$, while projecting $\boldsymbol{\varXi}$ back to the probability simplex can be performed in $\mathcal{O}(KU \log U)$ with a sorting-based routine. Line search evaluates the same objective structure for $I_{\text{LS}}$ times per iteration and thus costs $\mathcal{O}(KUM^2)$. Collecting all terms, the complexity per iteration is $\mathcal{O}( I_{\text{LS}}KUM^2+KUM^2+KU( M+N) +KU\log U+M+N )$, leading to total complexity in the order of $\mathcal{O}\left(I_{\text{RCG}} I_{\text{LS}}KUM^2\right)$, where $I_{\mathrm{RCG}}$ denotes the number of RCG iterations.

\textbf{Algorithm 2} summarizes the overall RALM with an inner RCG step, clipped multiplier updates, and conditionally grown penalty. Under standard regularity, including Lipschitz Riemannian gradients, a compact product manifold, and a Riemannian constraint qualification at the limit, which hold here in our setting, the proposed RALM behaves like its Euclidean counterpart. If each inner RCG produces an approximate stationary point of the current augmented Lagrangian and the inner tolerance is gradually tightened, the generated sequence is bounded and every accumulation point satisfies the Riemannian Karush-Kuhn-Tucker (KKT) conditions of the original problem; once the penalty is large enough as an exact-penalty regime, any feasible stationary point of the augmented Lagrangian is a KKT point of (P2.1). Therefore, whether or not Algorithm 2 converges depends on the tolerances for the subproblems, and the ability of the subsolver, i.e., Algorithm 1, to return a point that satisfies them. In practice, we tie the inner accuracy to the feasibility by decreasing $\epsilon_{\text{RCG}}^{(\ell)}$ as the constraint violation $\iota_k$ shrinks, and declare convergence when the subproblem updates fall below the tolerances $\zeta_{\min}$ and $\epsilon_{\min}$. Finally, we perform binary thresholding for $\boldsymbol{\varXi}$.

\begin{algorithm}[!t]
\caption{Overall RALM for solving (P2.1)}
\label{alg:ralm}
\begin{algorithmic}[1]
\REQUIRE Initial point $\boldsymbol{z}^{(0)}\in\mathcal{R}_{\boldsymbol{z}}$, multipliers $\boldsymbol{\lambda}^{(0)}\in\mathbb{R}^{K}$, subproblem tolerance $\epsilon_{\text{RCG}}^{(0)}$, minimum tolerance $\epsilon_{\min}$, penalty $\rho^{(0)}$, constants $\theta_{\epsilon}\!\in\!(0,1)$, $\theta_{\rho}\!>\!1$, bounds $\lambda^{\max}$, $\lambda^{\min}$ with $\lambda^{\min}\le\lambda^{\max}$, ratio $\varsigma_{\iota}\!\in\!(0,1)$, and minimum step $\zeta_{\min}$
\ENSURE $\boldsymbol{z}^{\star}$
\STATE $\ell\leftarrow 0$
\WHILE{$\mathrm{dist}(\boldsymbol{z}^{(\ell)},\boldsymbol{z}^{(\ell+1)})\geq \zeta_{\min}$ \OR $\epsilon_{\text{RCG}}^{(\ell)}\ge \epsilon_{\min}$}
    \STATE Obtain $\boldsymbol{z}^{(\ell+1)}$ by solving
           $\min_{\boldsymbol{z}\in\mathcal{R}_{\boldsymbol{z}}} \mathcal{L}_{\rho^{(\ell)}}(\boldsymbol{z}^{(\ell)},\boldsymbol{\lambda}^{(\ell)})$, i.e., (P2.2), via \textbf{Algorithm 1}
           with accuracy~$\epsilon^{(\ell)}$.
    \FOR{$i\in\mathcal{I}$}
        \STATE $\lambda^{(\ell+1)}_{k}\leftarrow
               \mathsf{Clip}_{[\lambda^{\min},\lambda^{\max}]}
               \bigl(\lambda_{k}^{(\ell)}+ \rho^{(\ell)} q_{k}(\boldsymbol{z}^{(\ell+1)})\bigr)$
        \STATE $\iota^{(\ell+1)}_{k}\leftarrow
               \max\bigl\{q_{k}(\boldsymbol{z}^{(\ell+1)}),-\lambda^{(\ell)}_{k}/\rho^{(\ell)}\bigr\}$
    \ENDFOR
    \STATE $\epsilon_{\text{RCG}}^{(\ell+1)}\leftarrow\max\{\epsilon_{\min},\theta_{\epsilon}\epsilon^{(\ell)}\}$
    \IF{$\ell\!=\!0$
        \OR
        $\max _{k\in \mathcal{K}}\left\{ \iota _{k}^{(\ell +1)} \right\} \! \le \! \varsigma _{\iota}\max_{k\in \mathcal{K}}\left\{ \iota _{k}^{(\ell)}\right\}$}
        \STATE $\rho^{(\ell+1)}\leftarrow\rho^{(\ell)}$
    \ELSE
        \STATE $\rho^{(\ell+1)}\leftarrow\theta_{\rho}\,\rho^{(\ell)}$
    \ENDIF
    \STATE $\ell\leftarrow \ell+1$
\ENDWHILE
\RETURN $\boldsymbol{z}^{\star}=\boldsymbol{z}^{(\ell)}$
\end{algorithmic}
\end{algorithm}

\vspace{-3pt}
\section{Heuristic Closed-Form MIS Design}
\vspace{-3pt}
In this section, we derive closed-form phase distribution and position-shift designs for MIS, achieving two-dimensional beam steering for wireless sensing, yet do not explicitly address interference suppression. For ease of derivation, here we assume that both MS1 and MS2 have a sufficiently large aperture $\mathcal{S}$, so that the overlapping area remains large, playing a decisive role for the beam pattern, and the non-overlapped elements of MS1 and its effects on the beam pattern are marginal and thus neglected. Following this simplification, let $(x,y)\in\mathbb{R}^2$ be the aperture coordinates of MS1 and denote the phase functions per layer by $\phi(x,y)$ for MS1 and $\theta(x,y)$ for MS2 with the same reference coordinate point. During operation, MS2 is repositioned by a small displacement vector $\boldsymbol{r}^T\triangleq (\Delta x,\Delta y)\in\mathbb{R}^2$ within the plane. For a narrowband plane wave propagation with wavenumber $\omega =2\pi/\lambda_c$, the equivalent phase realized by the MIS at a point $(x,y)$ is
\begin{align}
v(x,y;\boldsymbol{r}) &\triangleq \phi(x,y)+\theta(x-\Delta x,y-\Delta y)
\end{align}
Under the standard array factor model for LoS propagation, the far-field pattern in direction $(\vartheta,\psi)$ is 
\begin{align}
\!\!\!\mathcal{F}(\vartheta,\psi;\boldsymbol{r}) \!\!\propto \!\!\! \iint_{\mathcal{S}}{\! e^{jv( x,y;\boldsymbol{r} ) -j\omega( x\sin\! \psi \cos\! \vartheta +y\sin \!\psi \sin \!\vartheta)}\mathrm{d} x \mathrm{d} y}. \!\!\!\!
\end{align}
From (47), when $v(x,y;\boldsymbol{r})$ realizes a linear phase ramp whose gradient matches $(\omega\sin\! \psi \cos\! \vartheta,\omega\sin \!\psi \sin \!\vartheta)$, the beam steering to $(\vartheta,\psi)$ can be achieved [32]. 

Impose a sign-reversal pairing for the phase functions [33]
\begin{align}
\theta(x,y) = -\phi(x,y).
\end{align}
Substituting (48) into (46) and applying a first-order Taylor expansion for small displacements between two layers yields
\begin{subequations}
\begin{align}
\!v(x,y;\boldsymbol{r}) &= \phi(x,y) - \phi(x-\Delta x,y-\Delta y) 
\\
\!&=\boldsymbol{r}^T\cdot \nabla \phi(x,y) .
\end{align}
\end{subequations}
Hence, the equivalent phase equals the inner product between the displacement vector and the gradient of the phase functions. If $\nabla\phi(x,y)$ is affine in $(x,y)$, then $v(x,y;\boldsymbol{r})$ is a linear function w.r.t. $(x,y)$, which is exactly the steering condition. Therefore, a convenient choice is the quadratic template [34]
\begin{align}
\phi(x,y) &\triangleq A\big(x^2+y^2\big)+\phi _{\text{const}},\quad A>0,
\end{align}
with pre-determined constants $A$ and $\phi_{\text{const}}$. From (50) we have $\nabla\phi=[2Ax,~2Ay]^T$, and the equivalent phase becomes
\begin{equation}
v(x,y;\boldsymbol{r}) = 2A(\Delta x\cdot x +\Delta y\cdot y),
\end{equation}
which is a linear function with gradients being $\boldsymbol r$ w.r.t. $(x,y)$. Matching (47) with the array factor $(\omega\sin\! \psi \cos\! \vartheta,\omega\sin \!\psi \sin \!\vartheta)$ gives a closed-form MIS beam steering law to move MS2 as
\begin{subequations}
\begin{align}
&\Delta x=\frac{\omega }{2A}\sin \psi \cos \vartheta
\\
&\Delta y=\frac{\omega }{2A}\sin \psi \sin \vartheta
\end{align}
\end{subequations}
or, conversely, the beam steering angles can be recovered as
\begin{subequations}
\begin{align}
&\vartheta =\mathrm{arctan} \Big( \frac{2A}{\omega }\Delta y,\frac{2A}{\omega }\Delta x \Big), 
\\
&\psi =\mathrm{arcsin} \Big( \frac{2A}{\omega }\sqrt{\left( \Delta x \right) ^2+\left( \Delta y \right) ^2} \Big). 
\end{align}
\end{subequations}
The constant term $\phi_{\rm const}$ does not affect the steering, chosen aligned to the array response $\mathbf{a}_{\text{MIS}}$ in the BS direction. We can also find a design insight that $A$ trades coverage for resolution. Specifically, the larger $A$ broadens the angular coverage but coarsens the angular quantization for a given step $(\Delta x,\Delta y)$. 

To quantitatively design $A$, we derive the lower bound of the achievable angular coverage and the upper bound of the angular resolution. Let the maximum admissible displacements satisfy $|\Delta x|\le \Delta x^{\max}$ and $|\Delta y|\le \Delta y^{\max}$. The set of reachable directions, that is, the achievable angular coverage, $\vartheta \in \left[ \vartheta _{\min},\vartheta _{\max} \right] $ and $ \psi \in \left[ \psi _{\min}, \psi _{\max} \right]$ with -$\pi \le \vartheta _{\min}\le \vartheta _{\max}\le \pi $ and $0\le \psi _{\min}\le \psi _{\max}\le \pi$, respectively, satisfies
\begin{subequations}
\begin{align}
&\underset{\left( \vartheta ,\psi \right)}{\max}|\sin \psi \cos \vartheta |\le \frac{2A}{\omega }\Delta x_{\max},
\\
&\underset{\left( \vartheta ,\psi \right)}{\max}|\sin \psi \sin \vartheta |\le \frac{2A}{\omega }\Delta y_{\max}.
\end{align}
\end{subequations}
To make the above inequalities hold, we have
\begin{align}
\!\!A_{\text{cov}}\!=\!\frac{\omega}{2}\max \! \bigg\{ \frac{\underset{\left( \vartheta ,\psi \right)}{\max}|\sin \psi \cos \vartheta |}{\Delta x_{\max}},\frac{\underset{\left( \vartheta ,\psi \right)}{\max}|\sin \psi \sin \vartheta |}{\Delta y_{\max}} \bigg\}. \!\!
\end{align}
In addition, according to (54), for any given azimuth angle $\psi$ with a determined $A$, the maximum elevation angle $\vartheta$ satisfies
\begin{align}
\!\left| \psi _{\max}(\vartheta) \right|\!=\!\mathrm{arcsin}\! \left(\!\min \left\{\! 1,\frac{2A}{\omega}\frac{\Delta x_{\max}}{\left| \cos \vartheta \right|},\frac{2A}{\omega}\frac{\Delta y_{\max}}{\left| \sin \vartheta \right|} \!\right\} \!\right)\!, \!\!\!
\end{align}
where we find the principal planes: at $\psi=0$ or $\psi=\pi$, $|\psi| \le \mathrm{arcsin} \left( \frac{2A}{\omega }\Delta x_{\max} \right) $; at $\vartheta =\pm \frac{\pi}{2}$, $|\psi| \le \mathrm{arcsin}\sin \left( \frac{2A}{\omega }\Delta y_{\max} \right)$.

\setlength{\abovecaptionskip}{3pt}
\begin{figure*}[t]
    \centering
    \subfloat[Beam pattern 1 for target 1.]{
    \includegraphics[width=2.1in]{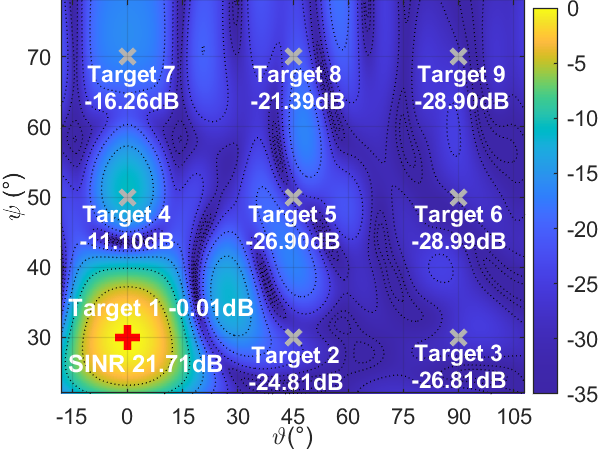}}
    \hspace{-0.05in}
    \subfloat[Beam pattern 2 for target 2.]{
        \includegraphics[width=2.1in]{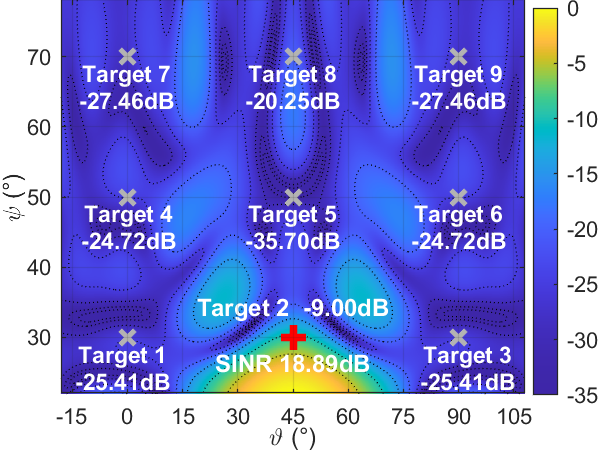}}
    \hspace{-0.05in}
    \subfloat[Beam pattern 3 for target 3.]{
        \includegraphics[width=2.1in]{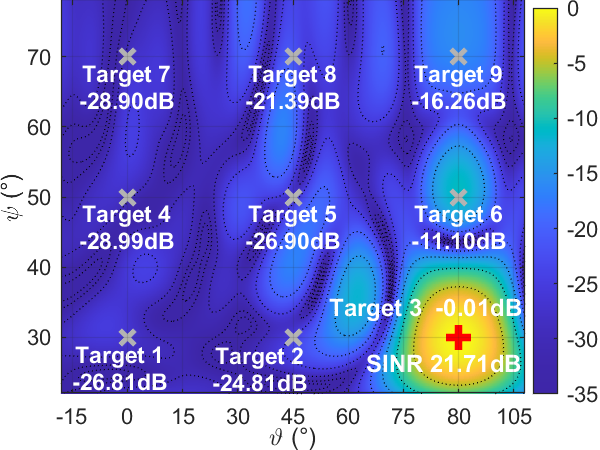}}
    \vspace{-6pt}
    \\
    \subfloat[Beam pattern 4 for target 4.]{
        \includegraphics[width=2.1in]{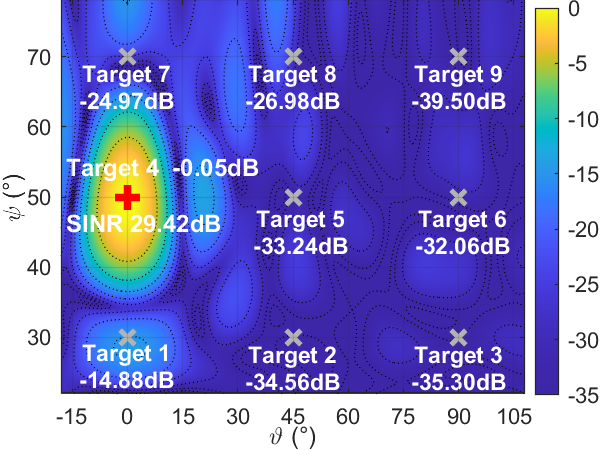}}
    \hspace{-0.05in}
    \subfloat[Beam pattern 5 for target 5.]{
        \includegraphics[width=2.1in]{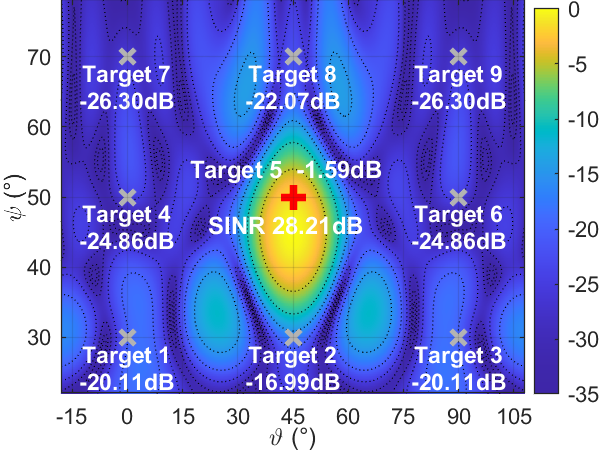}}
    \hspace{-0.05in}
    \subfloat[Beam pattern 6 for target 6.]{
        \includegraphics[width=2.1in]{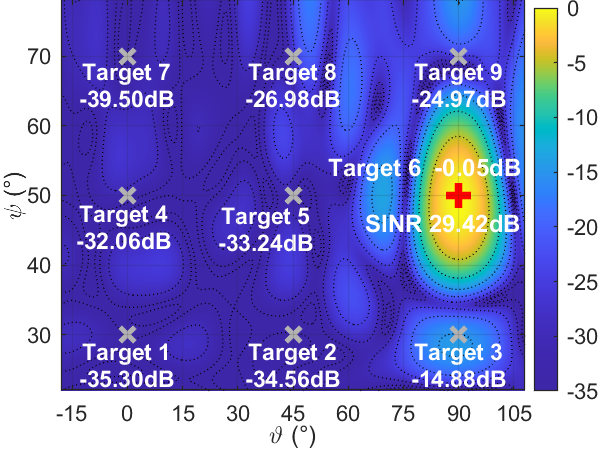}}
    \vspace{-6pt}
    \\
    \subfloat[Beam pattern 7 for target 7.]{
        \includegraphics[width=2.1in]{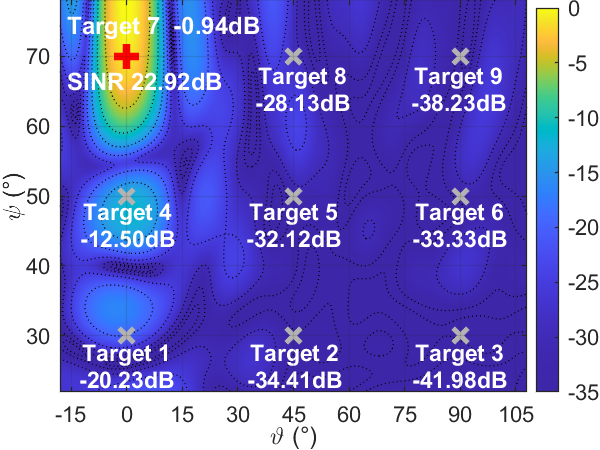}}
    \hspace{-0.05in}
    \subfloat[Beam pattern 8 for target 8.]{
        \includegraphics[width=2.1in]{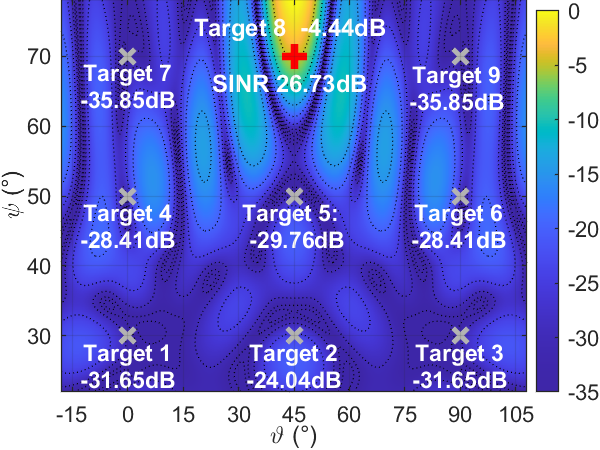}}
    \hspace{-0.05in}
    \subfloat[Beam pattern 9 for target 9.]{
        \includegraphics[width=2.1in]{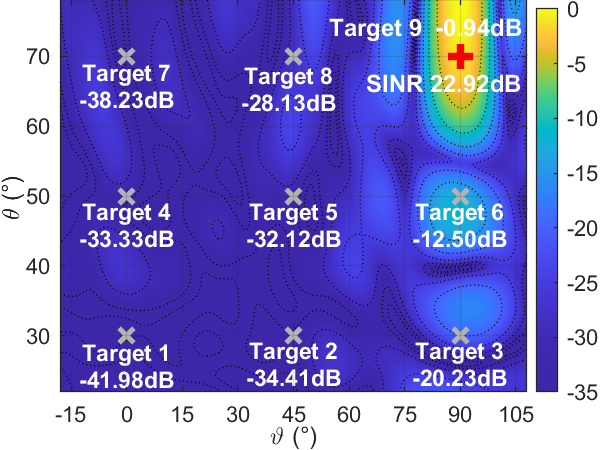}}
    \label{fig:beam_steering}
    \captionsetup{font=small}
    \caption{Normalized beam pattern gain and sensing SINR of the proposed MIS scheme with the closed-form design in Section V.} 
    \vspace{-15pt}
\end{figure*}

Further, to evaluate the angular resolution, we first derive the complete differential of $(\vartheta,\psi)$ w.r.t. $(\Delta x,\Delta y)$ according to the functional relationship in (53), 
\begin{subequations}
\begin{align}
\mathrm{d}\vartheta =&\frac{\Delta y\mathrm{d}( \Delta x ) +\Delta x \mathrm{d}( \Delta y )}{( \Delta x) ^2+( \Delta y ) ^2}
\\
=&\frac{2A}{\omega }\frac{-\sin \vartheta \mathrm{d}( \Delta x ) +\cos \vartheta \mathrm{d}( \Delta y )}{\sin \psi},
\end{align}
\vspace{-6pt}\begin{align}
\mathrm{d}\psi =&\frac{\mathrm{d}\big( \sqrt{( \Delta x ) ^2+( \Delta y) ^2} \big)}{\sqrt{1-( \Delta x ) ^2-( \Delta y ) ^2}}
\\
=&\frac{2A}{\omega }\frac{\cos \vartheta \mathrm{d}( \Delta x ) +\sin \vartheta \mathrm{d}( \Delta y )}{\cos \psi}.
\end{align}
\end{subequations}
Given a minimum MIS step size $(d_x,d_y)$, the worst-case grid-alignment requirement
$\left( \left| \Delta \psi _{\text{tar}} \right|,\left| \Delta \vartheta _{\text{tar}} \right| \right)$ over the target angle domain implies the upper bound of angular resolution
\begin{subequations}
\begin{align}
&\left| \Delta \vartheta _{\text{tar}} \right| \ge \left| \Delta \vartheta \right|= \frac{2A}{\omega }\frac{\left| \sin \vartheta d_x-\cos \vartheta d_y\right|}{\sin \psi},
\\
&\left| \Delta \psi _{\text{tar}} \right|\ge \left| \Delta \psi \right|=\frac{2A}{\omega }\frac{\left| \cos \vartheta d_x+\sin \vartheta d_y \right|}{\cos \psi}.
\end{align}
\end{subequations}
To guarantee worst-case beam alignment for any grid, we have
\begin{align}
\!\!A_{\text{res}}\!=\! \frac{\omega }{2}\underset{\left( \vartheta ,\psi \right)}{\min}\!\left\{\! \frac{\sin \psi \left| \Delta \vartheta _{\text{tar}} \right|}{\left| \sin \vartheta d_x\!-\!\cos \vartheta d_y \right|},\frac{\cos \psi \left| \Delta \psi _{\text{tar}} \right|}{\left| \cos \vartheta d_x\!+\!\sin \vartheta d_y \right|} \!\right\} \!. \!\!
\end{align}
The formula verifies intuition on the trade-off between resolution and coverage about $A$. 

To guarantee reachability, $A$ must satisfy the coverage lower bound $A\ge A_{\text{cov}}$; on the other hand, to avoid skipping target angles under finite step sizes $(d_x,d_y)$, $A$ should also try to obey the resolution upper bound $A\le A_{\text{res}}$. If $A_{\text{cov}}\le A_{\text{res}}$, we choose an optimal $A^*=A_{\text{res}}$ that maximizes the angular resolution. However, if the feasible set is empty due to $A_{\text{cov}}\ge A_{\text{res}}$, we need to choose $A^*=A_{\text{cov}}$ to guarantee coverage. In practice, we design a general coverage-guaranteed choice, based on MIS configurations $\Delta x_{\max}=(U_r\!-\!1)d_{\text{MIS}}$ and $\Delta y_{\max}=(U_c\!-\!1)d_{\text{MIS}}$, as $A\!=\!\frac{\pi}{\lambda_c d_{\text{MIS}}}\max\! \big\{\! \frac{1}{U_r-1},\frac{1}{U_c-1}\! \big\}\! \ge\! A_{\text{cov}}$, which suffices to achieve good angular resolution with a sufficiently large aperture size.

\setlength{\abovecaptionskip}{3pt}
\begin{figure*}[t]
    \centering
    \subfloat[Beam pattern 1 for target 1.]{
    \includegraphics[width=2.1in]{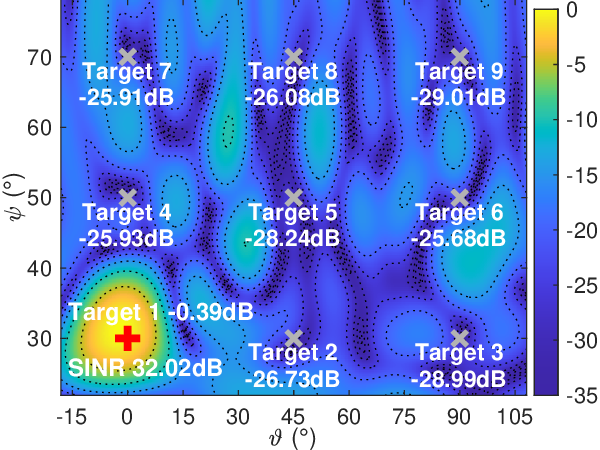}}
    \hspace{-0.05in}
    \subfloat[Beam pattern 2 for target 2.]{
        \includegraphics[width=2.1in]{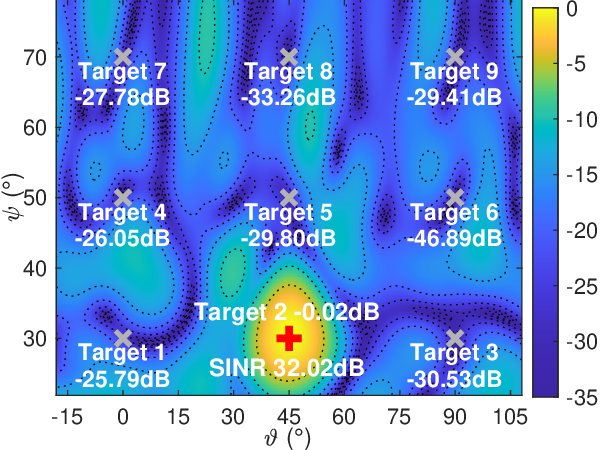}}
    \hspace{-0.05in}
    \subfloat[Beam pattern 3 for target 3.]{
        \includegraphics[width=2.1in]{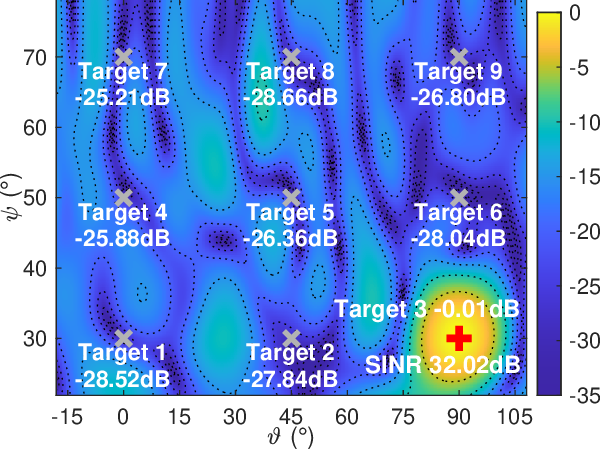}}
    \vspace{-6pt}
    \\
    \subfloat[Beam pattern 4 for target 4.]{
        \includegraphics[width=2.1in]{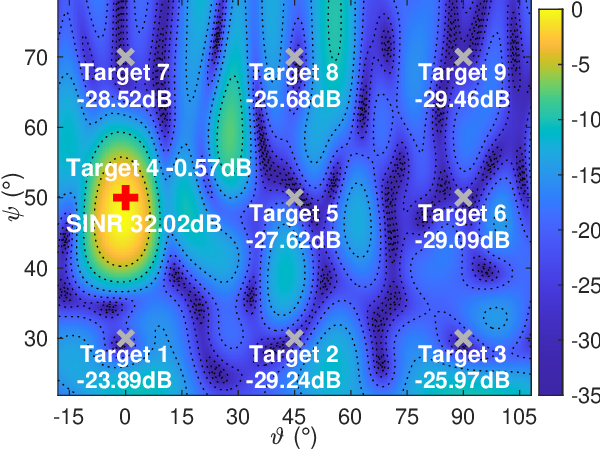}}
    \hspace{-0.05in}
    \subfloat[Beam pattern 5 for target 5.]{
        \includegraphics[width=2.1in]{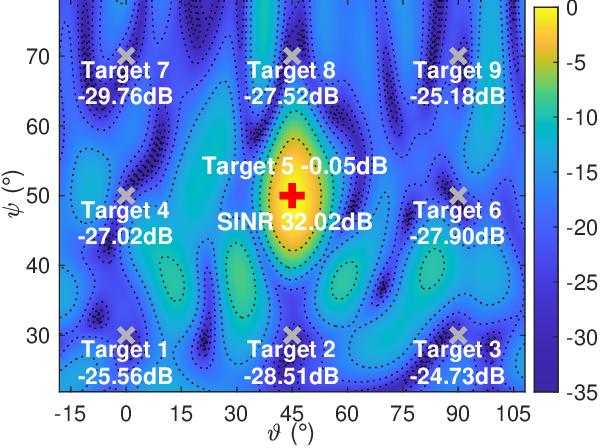}}
    \hspace{-0.05in}
    \subfloat[Beam pattern 6 for target 6.]{
        \includegraphics[width=2.1in]{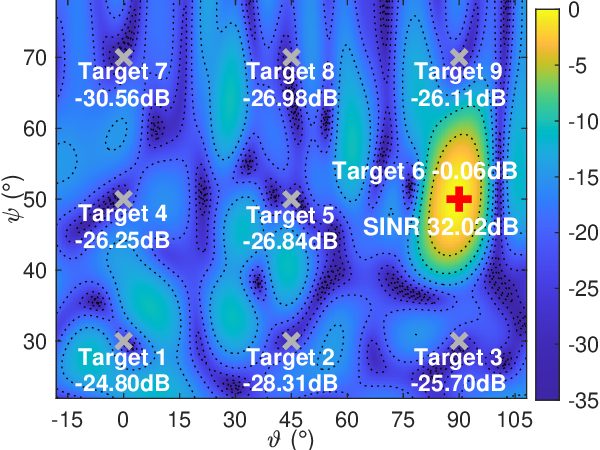}}
    \vspace{-6pt}
    \\
    \subfloat[Beam pattern 7 for target 7.]{
        \includegraphics[width=2.1in]{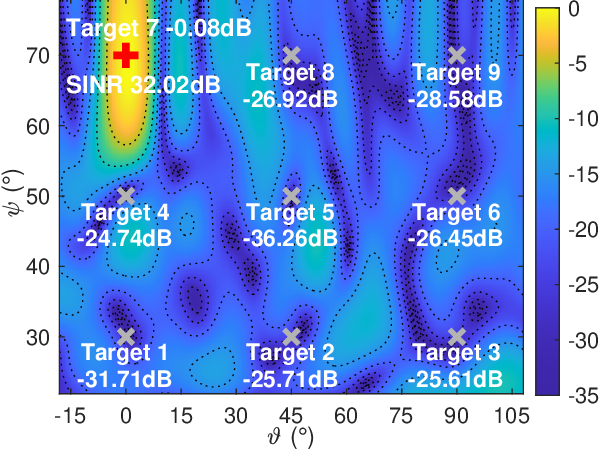}}
    \hspace{-0.05in}
    \subfloat[Beam pattern 8 for target 8.]{
        \includegraphics[width=2.1in]{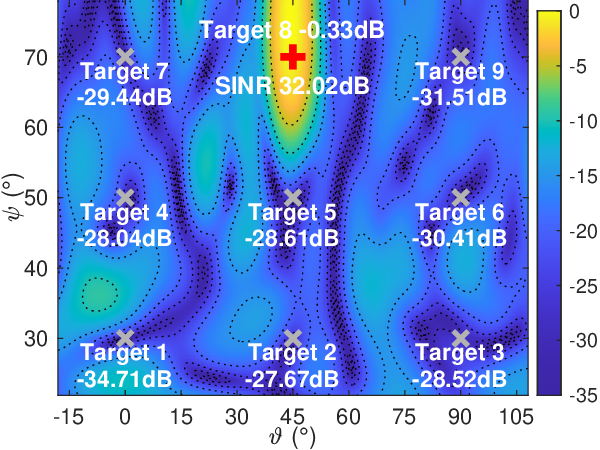}}
    \hspace{-0.05in}
    \subfloat[Beam pattern 9 for target 9.]{
        \includegraphics[width=2.1in]{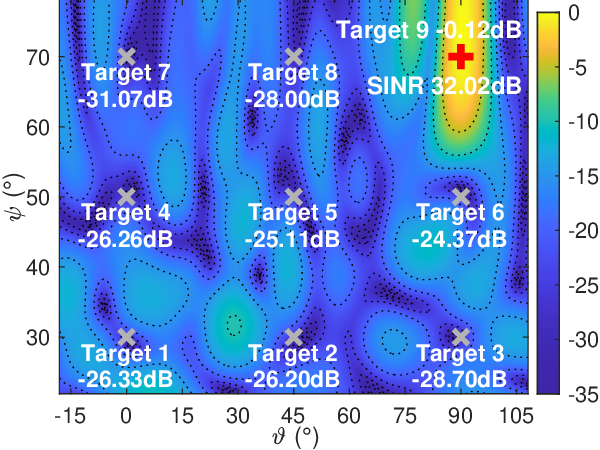}}
    \label{fig:beam_steering}
    \captionsetup{font=small}
    \caption{Normalized beam pattern gain and sensing SINR of the proposed MIS scheme with the optimization-based design in Section IV.} 
    \vspace{-15pt}
\end{figure*}

\vspace{-7pt}
\section{Numerical Results}
\vspace{-2pt}
This section evaluates the proposed MIS-enabled sensing schemes with both the heuristic closed-form design and the optimization-based RALM design. The BS is located on the surface normal of the MIS, while the targets are placed on the opposite side with uniformly distributed angles $\psi_k\in[30°,70°]$ and $\vartheta_k\in[0°,90°]$. The multi-hop path loss of the sensing links and the noise power are combined into a reference echo SNR of $\frac{\beta _{k}^{2}}{\sigma _{k}^{2}}=-73.88\text{dB}$ for all $k$. The carrier is $f_c=12$ GHz with wavelength $\lambda_c=0.025$ m, and the spacing of the MIS elements is $d_{\text{MIS}}=\frac{\lambda_c}{3}$. To isolate the effect of MIS control, we use $L=1$ BS antenna, which does not scale SINR in our model, and transmit power $P=30$ dBm. Unless stated otherwise, the desired elevation and azimuth grid sizes are $K_\psi=K_\vartheta=2$, and hence $K=K_\psi K_\vartheta=4$.

Figs. 2 and 3 plot the normalized beam pattern gains and annotate the resulting sensing SINR at each target. In the case study, we visualize nine discrete target directions with $K_\psi=K_\vartheta=3$. The MIS sizes are $M_r=M_c=20$ and $N_r=N_c=16$. The movable layer thus provides $U=U_r\times U_c=25$ overlapping positions, and nine of them are scheduled to illuminate the corresponding angular points. In Fig. 2, the closed-form scheme generates each beam by translating the movable MS2 according to the displacement law in Section V, without explicit interference suppression for the signal leakage to undesired target directions. 
As a result, beam centering is imperfect at some angles, and side-lobe leakage is only moderately suppressed, which may even exceed $-11$ dB. This yields a wide SINR spread across the nine targets: while favorable angles (e.g., targets 4 and 6) reach about $29.42$ dB, the worst case (target 2) drops to $\approx18.89$ dB due to off-center main-lobe gain and notable side-lobe leakage. Despite its simplicity and low complexity, the closed-form method therefore behaves like a time-division beam coverage mechanism: it can point toward different directions but offers limited fairness and weak leakage control when multiple sensing angles must be supported with a fixed hardware configuration. In contrast, the RALM-optimized design, shown in Fig. 3, explicitly maximizes the worst-case sensing SINR over all scheduled directions by jointly selecting the MIS position and designing the phase configuration. The synthesized beams accurately steer toward each designated $(\psi,\vartheta)$ with negligible deviation, and the side-lobe floor is consistently pushed below around $-25$ to $-30$ dB. Consequently, all nine targets achieve an essentially identical SINR $\approx32.02$ dB with near-optimal fairness. Compared with the closed-form baseline, the minimum SINR improves by about $13$ dB. This uniformity is valuable for reliable multi-target detection, where performance is dominated by the weakest target direction.

\begin{figure}[t]
\centering
\includegraphics[width=2.8in,height=2.1in]{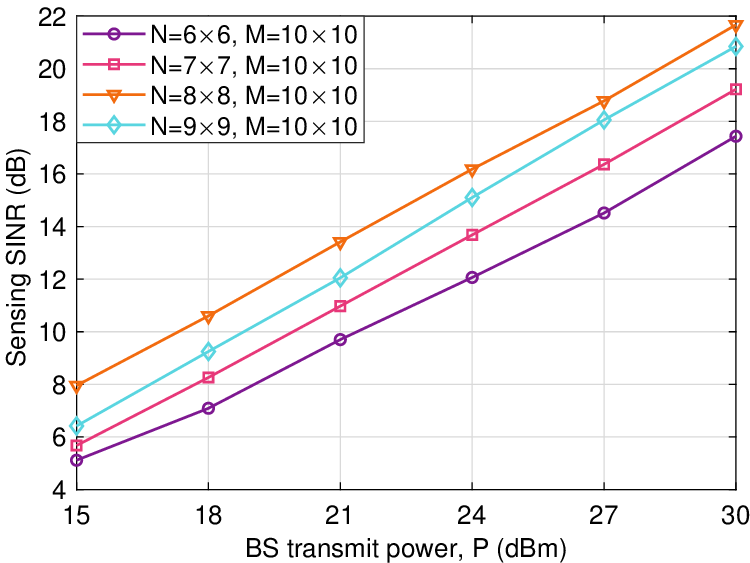}
\captionsetup{font=small}
\caption{Minimum SINR vs. transmit power for RALM-based design.} 
\vspace{-18pt}
\end{figure}

\begin{figure}[t]
\centering
\includegraphics[width=2.8in,height=2.1in]{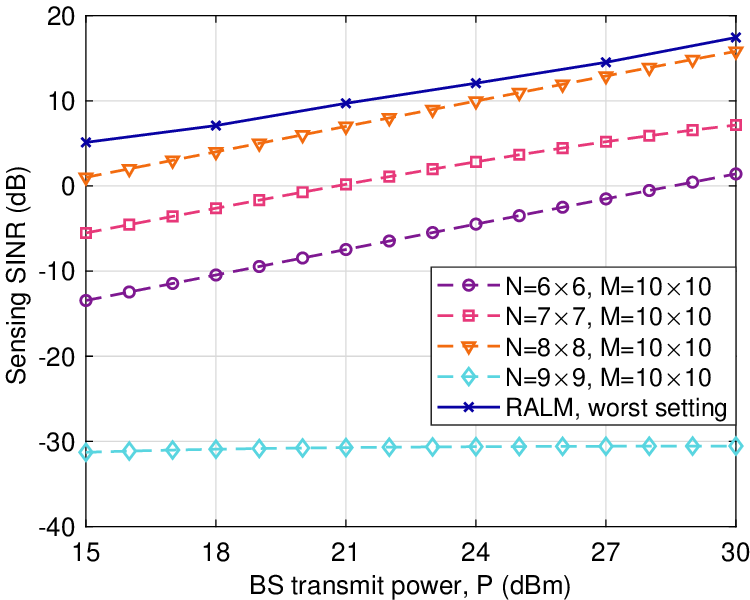}
\captionsetup{font=small}
\caption{Minimum SINR vs. transmit power for closed-form design.} 
\vspace{-18pt}
\end{figure}

Figs. 4 and 5 study the sensing SINR performance versus BS power $P$ and how the size of movable MS2 $N_r\times N_c$ affects sensing SINR under fixed $M_r\times M_c=10\times 10$ for MS1. Fig. 4 shows the RALM results versus BS power $P$ for $N_r=N_c\in\{6,7,8,9\}$. All curves increase almost linearly with $P$, indicating that inter-direction leakage is effectively suppressed by the optimization algorithm, as the system is essentially noise-limited. More importantly, performance is not monotonic in $N$. The best SINR is achieved around $N=8\times8$, followed closely by $N=9\times9$; reducing MS2 further to $N=7\times7$ and $6\times6$ degrades SINR. This reveals a key trade-off between effective pattern gain and pattern diversity: with position-shifting-based beam steering, the effective aperture of any reconfigurable beam pattern equals the overlap region between MS1 and the shifted MS2. A too small $N$ for MS2 limits its aperture and pattern gain, while a too large MS-2, with $N$ close to $M$, leaves insufficient position tuning flexibility. Sitting on each extreme can distort the pattern, lowering the gain on target and raising the sidelobes. Fig. 7 fixes the worst RALM configuration from Fig. 6 (the $N=6\times 6$ that gives the lowest SINR across powers) and contrasts it with closed-form designs using the same $N\in\{6,7,8,9\}$. The closed-form designs exhibit much poorer SINR even compared to this worst RALM curve, ranging from only a few dB in the best setting to deeply negative values, and for $N=9\times9$ remaining stuck around $-30$ dB, with almost no power gain, due to strictly constrained flexibility of mutual displacement.

\begin{figure}[t]
\centering
\includegraphics[width=2.8in,height=2.1in]{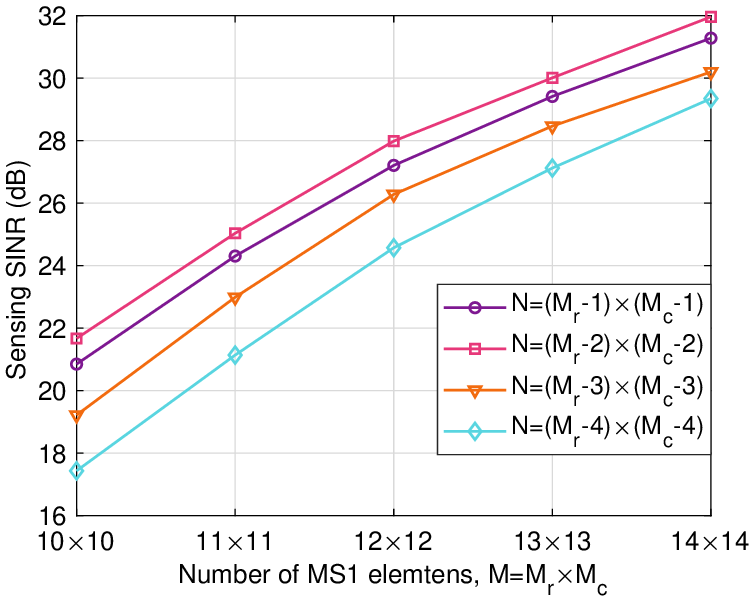}
\captionsetup{font=small}
\caption{Minimum sensing SINR vs. varying MIS sizes $(M,N)$ with fixed number of beam patterns $U$ for RALM-based design.} 
\vspace{-12pt}
\end{figure}

\begin{figure}[t]
\centering
\includegraphics[width=2.8in,height=2.1in]{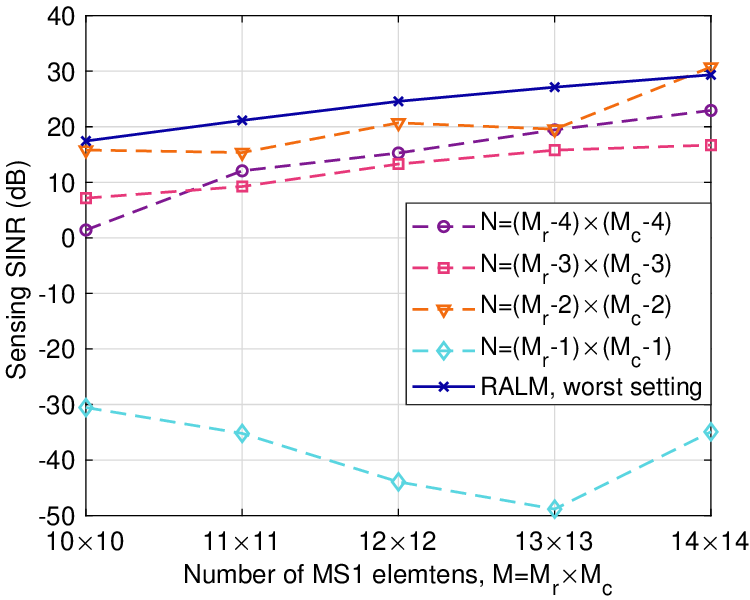}
\captionsetup{font=small}
\caption{Minimum sensing SINR vs. varying MIS sizes $(M,N)$ with fixed number of beam patterns $U$ for closed-form design.} 
\vspace{-18pt}
\end{figure}

Figs. 6 and 7 examine the regime where the beam pattern diversity is fixed while both MS1 and MS2 grow. We enforce the number of positions admissible for shifting $M_r-N_r=M_c-N_c\in\{1,2,3,4\}$ to be a constant. In Fig. 6, for every fixed size of the available beam pattern set $U=(M_r-N_r+1)\times(M_r-N_r+1)$, the sensing SINR increases consistently as $M$ grows from $10\times10$ to $14\times14$. This is expected: with diversity fixed, enlarging both panels primarily scales the effective aperture (i.e., overlap area) and, therefore, the gain on each desired target, while RALM keeps inter-direction leakage suppressed. Notably, the superiority of a moderate beam pattern diversity ($M_r-N_r=2$) over the other setting reaffirms the trade-off between pattern gain and diversity. Although $M_r-N_r=1$ provides the largest pattern gain of effective tunable aperture, it leaves too little edge room for position tuning to synthesis beam patterns with sufficient discrimination tailored for each target. In contrast, enlarging $M_r-N_r\in\{3,4\}$ overly shrinks the overlap area, losing the aperture gain, and suffering pattern aliasing across targets. Then, holding the same setting, Fig. 7 contrasts four closed-form baselines with the worst RALM curve in Fig. 6. The closed-form $M_r-N_r=1$ curve stays deeply negative ($\approx-30$ to $-35$ dB) across all $M$, indicating that sidelobes always overwhelm the desired echo. Closed-form $M_r-N_r\in\{2,3,4\}$ improves with $M$ but remains far below the RALM scheme due to the neglect of interference suppression. At large $M$, $M_r-N_r=2$ outperforms other closed-form settings and slightly exceeds the worst RALM curve, corroborating Fig. 6's insight that a moderate $U$ can be a near-optimal configuration.

\begin{figure}[t]
\centering
\includegraphics[width=2.8in,height=2.1in]{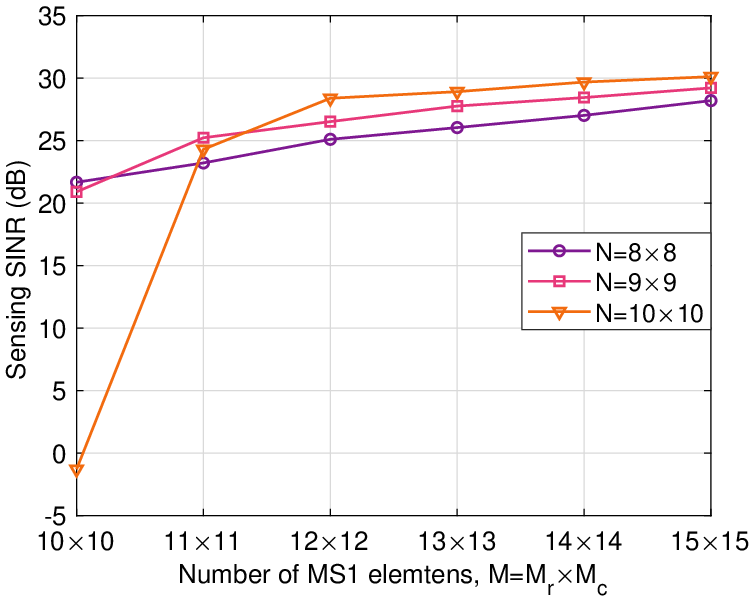}
\captionsetup{font=small}
\caption{Minimum sensing SINR vs. varying MS1 sizes $M$ with given MS2 sizes $N\in\{8\times 8, 9\times 9, 10\times10\}$ for RALM-based design.} 
\vspace{-12pt}
\end{figure}

\begin{figure}[t]
\centering
\includegraphics[width=2.8in,height=2.1in]{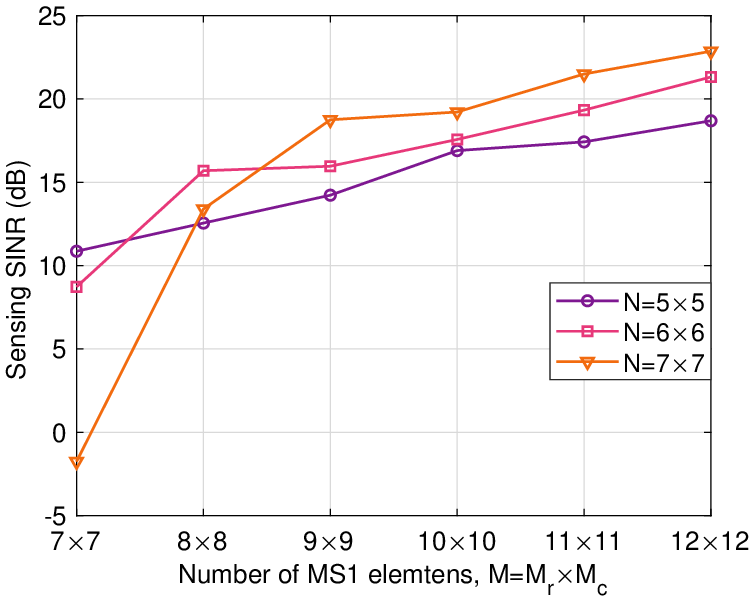}
\captionsetup{font=small}
\caption{Minimum sensing SINR vs. varying MS1 sizes $M$ with given MS2 sizes $N\in\{5\times 5,  6\times 6, 7\times7\}$ for RALM-based design.} 
\vspace{-15pt}
\end{figure} 

Figs. 8 and 9 study the RALM scheme when the size of movable MS2 $N_r\times N_c$ is fixed and the static MS1 $M_r\times M_c$ increases. The two figures use different $N$ sets but the same $M-N$ gap schedule, revealing the same mechanism under two operating ranges. When $M$ is only slightly larger than $N$, the system is limited by pattern diversity, as detailed in previous figures. This is most visible for the largest $N=7\times7$ in Fig. 8, as well as $N=10\times10$ in Fig. 9, at the smallest $M$, which even causes unusable performance. However, as $M_r$ and $M_c$ grow by just 1 to 2 elements per side, the scheme rapidly gains enough positional degrees of freedom, while the slope becomes the steepest among the three curves, as its effective tunable aperture begins to dominate. In contrast, the smallest MS2 setting $N=8\times8$ in Fig. 8, as well as $N=5\times5$ in Fig. 9, starts with a comfortable pattern diversity, due to larger $M_r-N_r$ and $M_c-N_c$, and thus performs reasonably well at small $M$, while its effective aperture is limited, so its growth with $M$ is the slowest. Following the above observations, the medium case such as $N=9\times9$ in Fig. 8 lies between these two extremes in both the starting level and in slope.

\begin{figure}[t]
\centering
\includegraphics[width=3in,height=2.2in]{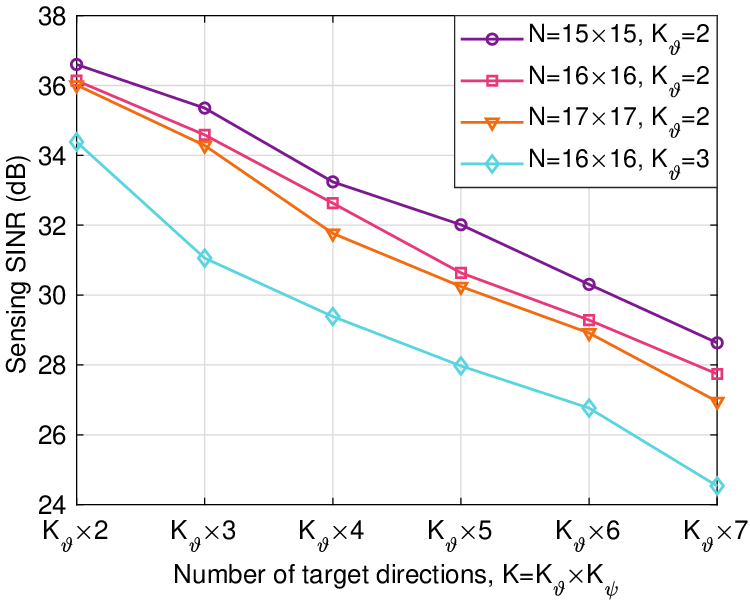}
\captionsetup{font=small}
\caption{Minimum sensing SINR vs. varying number of target directions $K$ for RALM-based design.} 
\vspace{-12pt}
\end{figure}

\begin{figure}[t]
\centering
\includegraphics[width=3in,height=2.2in]{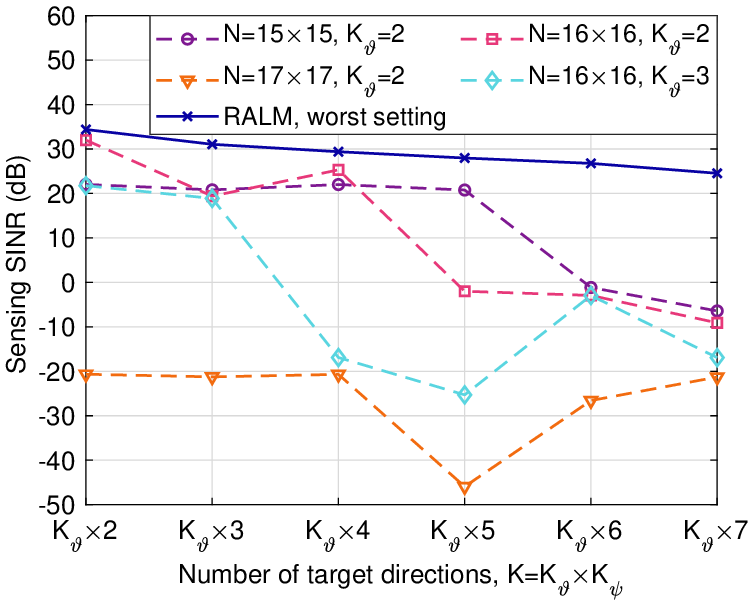}
\captionsetup{font=small}
\caption{Minimum sensing SINR vs. $K$ for closed-form design.} 
\vspace{-18pt}
\end{figure}

Figs. 10 and 11 show the impact of the number of target directions on the worst-case sensing SINR performance. RALM-based schemes with $M=20\times 20$ and different $N$ are shown in Fig. 10. As the number of target directions $K=K_\vartheta K_\psi$ grows, the minimum sensing SINR decreases due to stringent inter-direction interference suppression across more angles. For $K_\vartheta=2$, a larger movable layer yields a uniformly higher curve over all $K$ with a roughly constant gap. Keeping $N=16\times16$ and tightening the elevation grid to $K_\vartheta=3$ shifts the entire curve downward due to the stricter multi-direction sensing requirement. In Fig. 11, with the same system settings, the performance of the closed-form scheme either stays negative or deteriorates as $K$ increases, and sometimes shows non-monotonicity. In contrast, RALM curve among these settings declines only mildly, revealing the scalability of the proposed RALM approach for MIS-enabled multi-target sensing.

\section{Conclusion}
This paper proposed a low-cost MIS for wireless sensing, where a larger fixed MS1 and a smaller movable MS2 with pre-designed phase elements realize beam pattern reconfiguration via differential position shifting, without per-element tuning. Building on this architecture, we established a MIS-enabled multi-hop echo signal model in the presence of multi-target interference. We further formulated a worst-case sensing SINR maximization problem that jointly designs the MIS phase shifts and schedules MS2’s position. To solve the mixed-integer non-convex program, we developed an efficient RALM-based algorithm on a product manifold and, for a lightweight baseline, derived a heuristic MIS beam steering scheme with closed-form phase distribution and position-shifting expressions that enable two-dimensional beam scanning.
The numerical results validate the MIS’s beam pattern reconfiguration capability, demonstrate the superiority of the RALM-based scheme over the closed-form scheme for interference suppression and worst-case sensing SINR improvement, and reveal a clear trade-off between beam pattern gain and diversity. These observations provide practical guidance for selecting the MIS configuration parameters. Future research directions include wireless sensing with user mobility-aware MIS operation and hybrid MIS designs that combine static and tunable elements. 
In summary, MIS offers a cost-effective solution delivering wireless sensing capabilities, well-suited to sustainable 6G infrastructures.

\ifCLASSOPTIONcaptionsoff
  \newpage
\fi



%\bibliographystyle{IEEEtran}

\end{document}